\documentclass[aps,prb,floats,twocolumn]{revtex4-1}

\usepackage{
  amsmath,
  amssymb,
  array,
  booktabs,
  braket,
  color,
  float,
  graphicx,
  scrextend,
  hyperref,ulem
}

\usepackage[dvipsnames]{xcolor}
\usepackage[caption=false]{subfig}
\date{April 30, 2020}
\begin{document}

\title{Reactivity of transition-metal alloys to oxygen and sulphur}
\author{Rajarshi Tiwari}
\affiliation{School of Physics, AMBER and CRANN Institute, Trinity College Dublin, Dublin 2, Ireland}
\author{James Nelson}
\affiliation{School of Physics, AMBER and CRANN Institute, Trinity College Dublin, Dublin 2, Ireland}
\author{Chen Xu}
\affiliation{Nokia Bell Labs, 600 Mountain Avenue, Murray Hill, NJ, USA}
\author{Stefano Sanvito}
\affiliation{School of Physics, AMBER and CRANN Institute, Trinity College Dublin, Dublin 2, Ireland.}

\begin{abstract}
Oxidation and tarnishing are the two most common initial steps in the corrosive process of metals at ambient conditions. 
These are always initiated with O and S binding to a metallic surface, so that one can use the binding energy as a rough 
proxy for the metal reactivity. With this in mind, we present a systematic study of the binding energy of O and S across the 
entire transition-metals composition space, namely we explore the binding energy of {\bf 88} single-phase transition metals
and of {\bf 646} transition-metal binary alloys. The analysis is performed by defining a suitable descriptor for the binding 
energy. This is here obtained by fitting several schemes, based on the original Newns-Anderson model, against 
density-functional-theory data for the 4$d$ transition metal series. Such descriptor is then applied to a vast database of 
electronic structures of transition-metal alloys, for which we are able to predict the range of binding energies across both the 
compositional and the structural space. Finally, we extend our analysis to ternary transition-metal alloys and identify 
the most resilient compounds to O and S binding.
\end{abstract}

%\date{}
\maketitle
%\section{}
%\subsection}{

\section{Introduction}

The electronic industry uses a wide palette of metals in various forms. Tiny metallic wires form interconnectors in logic circuits, 
thin magnetic films are the media in data storage, mesoscopic layers are found as solders and protective finishings in printed 
circuit boards. All these metals undergo corrosion processes, which can lead to degradation and ultimately to failure. In the last 
few years the problem has aggravated because of the increased multiplicity of the elements used, the reduced spacing between 
the various components, the often unpredictable users' environment and the deterioration of the air quality in region with a high 
level of industrial activity. Thus, it is desirable to identify classes of metallic alloys, which are particular resilient to corrosion and, 
at the same time, can deliver the functionality desired by the given application.

There are several known mechanisms of corrosion depending on the environmental conditions, such as the mixture of corrosive 
agents at play and the humidity level, but a full experimental determination of the dynamics of corrosion is often difficult to achieve. 
In general, a corrosive reaction is initiated by the binding of a chemical agent to a metallic surface, followed by the formation of
a new phase, with or without the possible release of new molecules incorporating atoms from the metallic surface (mass loss).
The progression of the corrosive reaction beyond the formation of a few reacted atomic layers is then determined by the diffusion 
of the corrosive agent in the metal and the self-diffusion of the metallic ions. At the macroscopic level, such mass transport 
mechanisms are further determined by the microstructure of the given sample, for instance through diffusion at grain boundaries. 

Given the general complexity of the corrosion process modelling studies must extend across different length and time 
scales~\cite{Gunasegaram2014}. These studies usually provide a valid contribution to the understanding of the corrosion dynamics
in a given material. The multiscale approach, however, is not suitable for scanning across materials libraries in the search for the 
ideal compound resisting corrosion in a known environment, since the numerical overheads and workflows are computational 
prohibitive and often require information from experiments (e.g. the microstructure). Thus, if one wants to determine a simple set 
of rules to navigate the large chemical and structural space of metallic alloys, the attention has to focus on one of the steps 
encountered in the corrosion path. A natural choice is that of determining the ease with which the first step takes place, namely to 
evaluate the reactivity of a given metal to a particular chemical agent. 

This is precisely the approach adopted here, where we estimate the reactivity of a vast database of metallic alloys to both O and
S. Oxygen and sulphur are particularly relevant, since for many metallic surfaces oxidation and tarnishing initiate the corrosion 
process at the ambient conditions where one typically finds electronic equipment. However, even the simulation of oxidation and 
tarnishing is complex and not amenable to a high throughput study. This, in fact, involves determining the full reaction path through
an extensive scan of the potential energy surface or through molecular dynamics. As such here we take a simplified approach 
by computing the oxygen and sulphur binding energy to metals and by taking such binding energy as a proxy for reactivity. Clearly
this is a drastic simplification, since sometime a material presents a similar binding energy to O and S but different reactivity, as
in the case of silver~\cite{Saleh2019}. 
{Such situations typically occur when the rate-limiting barrier in the reaction path does not correlate well with the
binding energy of the reaction product (see discussion in section~\ref{sec:be-bin-o}), or when the interaction between 
the reactants on the surface changes the thermodynamics of the reaction as the coverage increases.}
However, the binding energy still provides a measure of the tendency of O and S to attack
the surface, and it strongly correlates to the stability of the product oxide/sulphide phases (see later). As such it is a valuable
quantity to classify metallic alloys as either weak or robust to corrosion.

Our computational scheme achieves the desired throughput by combining density functional theory (DFT) calculations with  an 
appropriate {\it descriptor}~\cite{Curtarolo2013}. In particular we use details of the density of state (DOS), namely the shape of 
the partial DOS (PDOS) associated to the transition metal $d$-bands, to construct a model that provides an estimate of the binding 
energy between a given metal alloy and both O and S. This is based on the notion that the O-metal and S-metal bonds get weaker 
as the metal $d$ band is progressively filled~\cite{Norskov1995}. Our strategy then proceeds as follows. We first fit the parameter of the
model to DFT binding energy data for the 4$d$ transition metal series. This is preferred to the 3$d$ one, since it does not present 
elemental phases with magnetic order, and to the 5$d$, since the electronic structure can be computed without considering spin-orbit 
interaction. In particular, we explore several variations of the model and assess their ability to fit the data. Then, we construct an 
automatic workflow to extract the DOS information from the AFLOWLIB.org materials database~\cite{Curtarolo2012}. This involves fitting
the various orbital-resolved PDOS to a semi-elliptical DOS. Finally, we use the descriptor to compute the binding energy for all the 
experimentally known binary and ternary transition-metal alloys contained simultaneously in both AFLOWLIB.org and the Inorganic Crystal
Structure Database (ICSD)~\cite{Zagorac2019}.

The paper is organized as follows. In the next section we introduce our methodology, by discussing the various descriptors 
considered, their rationale, the computed DFT data and the scheme for extracting data from AFLOWLIB.org. Next we present 
the descriptor fitting procedure and evaluate its error in determining the binding energy, before proceeding to show
our results. We first determine the binding energy of O and S to transition-metal binary alloys and then we move to a restricted
number of ternaries. Then we conclude.

\section{Methods}
\label{sec:meth}
\subsection{Rationale for the descriptors}

\begin{figure}[htb] 
\begin{center}
\includegraphics[width=0.45\textwidth]{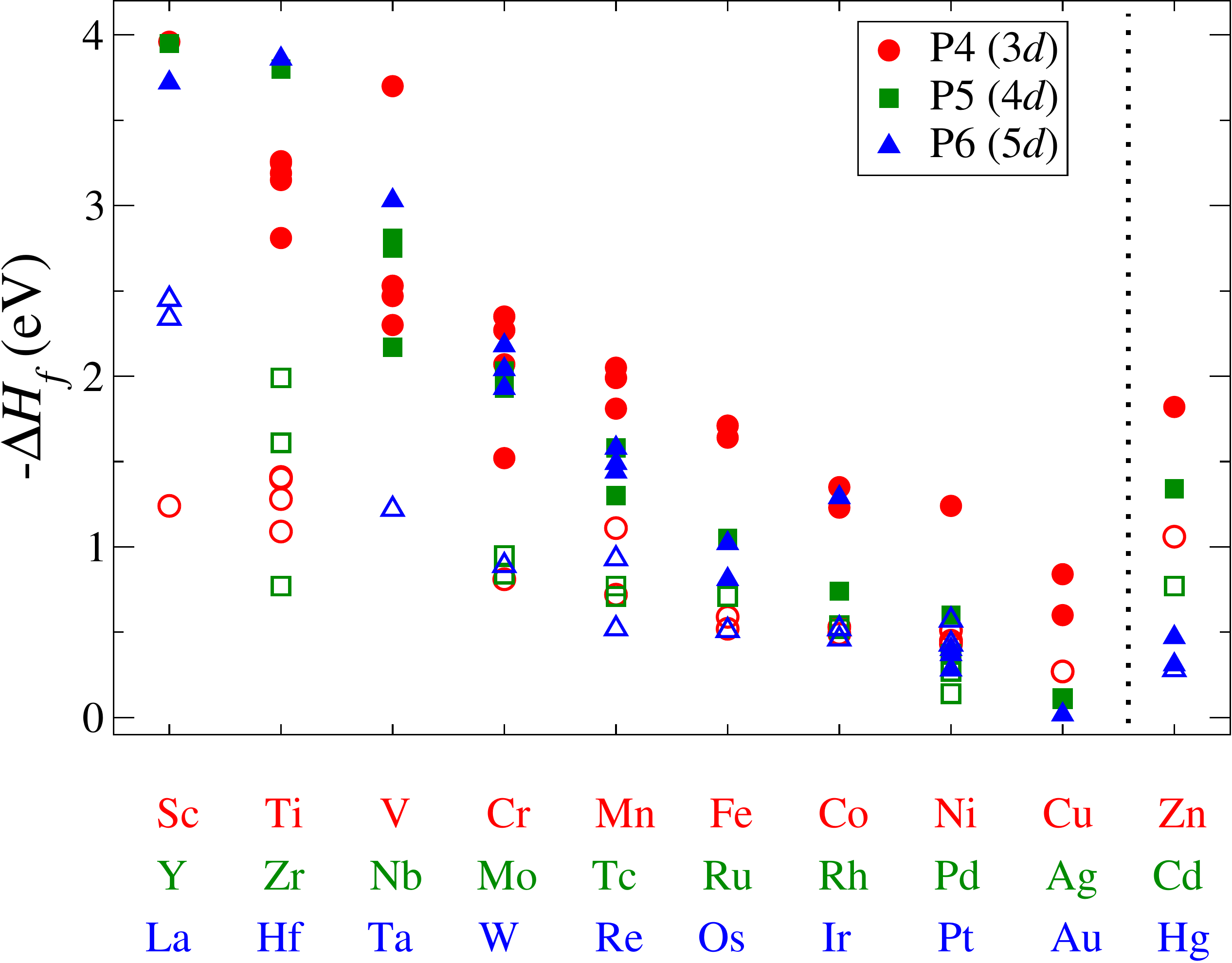}
\end{center}
\caption{(Color online) Experimental enthalpy of formation per atom for transition-metal oxides and sulphides across the 3$d$ (red 
circles), 4$d$ (green squares) and 5$d$ (blue triangles) series. Open symbols are for sulphides and closed symbols for 
oxides. Multiple symbols for the same transition metal correspond to oxides/sulphides with different stoichiometry. For 
instance, there are five different data points for Ti-O, respectively TiO, Ti$_2$O$_3$, Ti$_3$O$_5$ and TiO$_2$ (both
anatase and rutile). Note that for several transition metals additional stable phases exist,  but they are not reported in 
the figure because their enthalpy of formation is not available (e.g. for Ti-O there also exist Ti$_6$O, Ti$_3$O, Ti$_2$O,
and Ti$_3$O$_2$).}
\label{fig:1}
\end{figure}

The main idea beyond the definition of a descriptor is that it should represent a simple relation between a physical observable 
and a property of the electronic structure easy to calculate. Once this is established, the descriptor can be used to
scan large databases in the search for particular compounds of interest. In our case an insight on how to construct
a descriptor for the binding energy between O/S and a transition metal alloy can be obtained by looking at Fig.~\ref{fig:1}. 
In the figure we report the experimental enthalpy of formation per atom, $-\Delta H_f$, for a wide range of oxides and sulphides across 
the 3$d$, 4$d$ and 5$d$ transition-metal series (the data used for constructing Fig.~\ref{fig:1} are listed in the appendix together with
their associated references), where multiple data corresponding to the same transition metal indicate that oxides/sulphides with 
different stoichiometry exist for that metal.

The figure reveals a number of clear trends. Firstly,  we note that on average the enthalpies of formation of oxides are 
significantly larger than those of sulphides, owning to the fact that the electronegativity of O is larger than that of S. 
Secondly, for both oxides and sulphides the absolute value of the enthalpy of formation reduces monotonically (becomes 
less negative) across the transition-metal series. The slope of such reduction is significantly more pronounced for oxides
than for sulphides, so that towards the end of the series $-\Delta H_f$ is very similar for these two groups ($-\Delta H_f$ 
for Ag$_2$O is almost identical to that of Ag$_2$S, $\sim$0.11~eV/fu {- fu = formula unit}). Finally, the 
enthalpy of formation increases again beyond the noble metals (Cu, Ag and Au). 

Importantly, these trends have not been observed only for the enthalpy of formation but also for the binding energy of transition 
metals with monovalent atoms~\cite{Norskov1995,Hammer1995} (e.g. H), with oxygen~\cite{Besenbacher1993} and with a broad 
range of molecular adsorbates~\cite{Hammer1996,Rubam1997}. This suggests the formulation of a descriptor, characteristic of 
each adsorbate, based solely on the details of the electronic structure of the metal~\cite{Norskov1995}. The crucial observation is 
that in typical transition metals the DOS is dominated by a partially filled, extremely broad $s$-$p$ band, and by a relatively narrow 
$d$ band. As the atomic number increases the occupation of the $s$-$p$ band changes little, while the $d$ band becomes 
progressively more filled, and hence moves to lower energies (with respect to the Fermi energy, $E_\mathrm{F}$). Upon approaching 
the surface, the energy level of the adsorbate relevant for the bonding [in the case of O (S) the 2$p$ (3$p$) shell] gets broadened by the 
interaction with the $s$-$p$ band. At the same time, it forms a bonding and anti-bonding pair with the $d$ band of the metal, which is here 
approximated as a single molecular level. Thus, at the beginning of the transition-metal series one has the situation in which the 
bonding level is filled and the anti-bonding one is empty, so that the binding energy is high. However, as the $d$ band fills also the 
occupation of anti-bonding level increases, the adsorbate-metal bond weakens and the binding energy reduces. 

It is then clear that the energy position of the $d$ band of the transition metal, together with some measure of the strength of the 
transition-metal/adsorbate hopping parameter, can define a valuable descriptor for the binding energy. Two classes of such 
descriptors are defined in the next section.

\subsection{Definition of the descriptors}
\label{sec:descr}

We model oxygen and sulfur as a single impurity level coupled to a bath of electrons characterising the metal. The level of 
description for such bath defines the different models.
\begin{figure}[t]
  \centering
  \includegraphics[width=8cm, height=3cm]{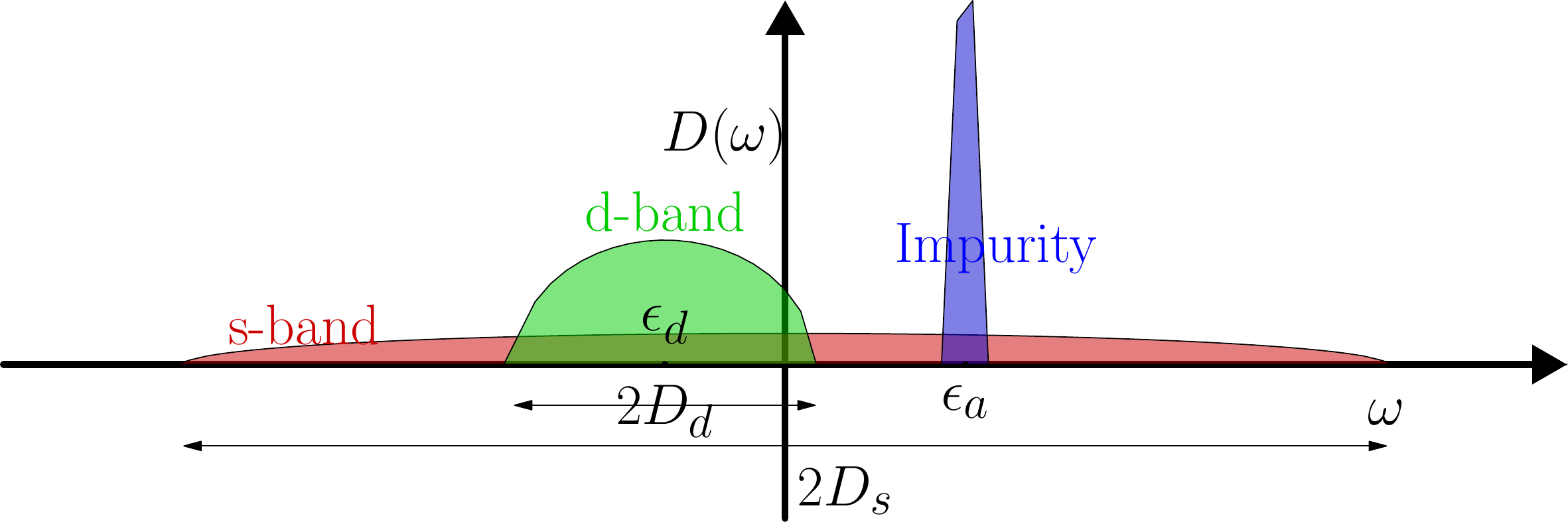}
  \caption{(Color online) Level scheme showing the DOS for the Anderson impurity model. The $s$-$p$ band of the metal is wide and 
  the Fermi level is placed approximately at half filling. In contrast, the $d$ band has a moderate width and it is centered 
  at $\epsilon_d$. The impurity level, whose width is determined only by the interaction with the metal, is at an energy, $\epsilon_a$. 
  }
  \label{fig:levels}
\end{figure}
The simplest one is often called the Newns-Anderson (NA) model~\cite{PhysRev.178.1123}, and it is defined by the following
Hamiltonian, 
\begin{equation}
  \label{eq:ham-na}
  H_\mathrm{NA} = \epsilon_d d^{\dagger} d + \epsilon_a a^{\dagger} a + V_{ad} (a^{\dagger} d + d^{\dagger} a)\:.
\end{equation}
Here $a^\dagger$ ($a$) and $d^\dagger$ ($d$) are the creation (annihilation) operators, respectively, for the adsorbate and 
the metal $d$ band, while $\epsilon_a$ and $\epsilon_d$ are the corresponding energy levels (before binding), and $V_{ad}$
their hybridization (hopping integral). At this level the metal $d$ band is treated as dispersionless and the contribution from the
$s$-$p$ band to the bond is neglected. $H_\mathrm{NA}$ can be easily diagonalised to yield the eigenvalues 
$\epsilon_{\pm} = \epsilon_a + \frac 12 (\Delta_{ad} \pm W_{ad})$, where, $\Delta_{ad} = \epsilon_d - \epsilon_a$ and 
$W_{ad} = \sqrt{4V^2_{ad}+\Delta^2_{ad}}$. If the fractional occupation of the $d$ band is $f$, then the total energies
before ($V_{ad}=0$) and after ($V_{ad}\ne0$) the coupling are $E_1 = 2(f\epsilon_d + \epsilon_a)$ and $E_2 = 2(1+f)\epsilon_{-}$,
respectively. Thus, the binding energy is given by
\begin{equation}
  \label{eq:be-na}
  E_\mathrm{b} = E_2 - E_1 = - (1-f)(W_{ad} - \Delta_{ad})\:.
\end{equation}
Finally, one can assume that there is an additional contribution to the binding energy, $E_{sp}$, originating from the 
interaction with the $s$-$p$ band. Such contribution, however, is not expected to vary much across the transition-metal 
series so that it can be kept constant. The NA model, thus depends on four parameters. Two of them are associated 
respectively, to the metal, $\epsilon_d$, and the adsorbate, $\epsilon_a$, alone, one to the interaction between the two, 
$V_{ad}$, and one is a constant, $E_{sp}$, specific of each adsorbate. 
{Note that when extracting the various parameters from electronic structure theory calculations (see next section),
where the $d$ band has dispersion, the $d$-band energy level, $\epsilon_d$, is replaced by the position of the band
center.}

A more detailed description of the electrons in the metal is provided by the Anderson impurity model,~\cite{PhysRev.124.41}
which is schematically illustrated in Fig.~\ref{fig:levels}. In this case the metal band structure is taken into consideration
through the Hamiltonian,
\begin{align}\nonumber
  \label{eq:Ham}
  H_\mathrm{A} &= {\epsilon_a a^{\dagger}a}
      + \sum_{{\bf k}}\left({\epsilon^s_{\bf k}s^{\dagger}_{\bf k}s_{\bf k} + \epsilon^d_{\bf k}d^{\dagger}_{\bf k}d_{\bf k}}\right)+\\
    &+ \sum_{{\bf k}}\left(V^s_{\bf k}a^{\dagger}s_{\bf k} + V^d_{\bf k}a^{\dagger}d_{\bf k} + \textrm{h.c.}\right)\:,
\end{align}
where now the operators $s^{\dagger}_{\bf k} (s_{\bf k}$) and $d^{\dagger}_{\bf k} (d_{\bf k}$) create (destroy) an electron
with wave-number ${\bf k}$, respectively, in the $s$-$p$ and in the $d$ band. The band energies are $\epsilon^s_{\bf k}$ 
and $\epsilon^d_{\bf k}$ and the hopping parameters $V^s_{\bf k}$ and $V^d_{\bf k}$. 

The model defined by Eq.~(\ref{eq:Ham}) can be solved by constructing the appropriate Green's function, as shown in detail 
in Appendix~\ref{appdx-1}. In brief, the `impurity' Green's function can be written as
\begin{equation}
G_{aa}(\omega) = {1 \over \omega -\epsilon_a - \Sigma(\omega)}\:,
\end{equation}
where $\Sigma(\omega)$ is the self energy describing the interaction with the metal. This is given by
\begin{align}
  \Sigma(\omega) = \sum_{\bf k}\left( \frac{|V^d_{\bf k}|^2}{\omega -\epsilon^d_{\bf k} + i\eta} + \frac{|V^s_{\bf k}|^2}{\omega -\epsilon^s_{\bf k} + i\eta} \right)\:,
\end{align}
with $\eta\rightarrow0^+$. If we assume that the couplings are independent of ${\bf k}$, namely $V^d_{\bf k}=V_d$ and $V^s_{\bf k}=V_s$,
we can simplify the self energy into $\Sigma(\omega) = \Lambda(\omega) - i\pi\Delta(\omega)$, so that the DOS, 
$D_a(\omega) = -\frac 1\pi\text{Im}[G_{aa}(\omega)]$, writes as
\begin{align}\label{DOS-imp}
  D_a(\omega) = { \Delta(\omega) \over [\omega - \epsilon_a - \Lambda(\omega)]^2 + \pi^2\Delta(\omega)^2 }\:.
\end{align}
Explicit expressions for the real, $\Lambda(\omega)$, and imaginary part, $\Delta(\omega)$, of the DOS are detailed in Appendix \ref{appdx-1}.
Finally, the binding energy can be obtained by integrating the DOS, 
\begin{align}
  \label{eq:be-2}
  E_\mathrm{b} = \int_{-\infty}^{0}D_a(\omega)d\omega - \epsilon_a\:,
\end{align}
where all the energies are defined from the metal Fermi energy, $E_\mathrm{F}=0$.

As defined, the Anderson impurity model depends on the adsorbate energy level, the metal/adsorbate electronic 
coupling and the metal DOS. Here we approximate the metal DOS with a semi-circular model. In particular the
$d$-band DOS, $D_d(\omega)$, is described as a semi-circle with center at $\epsilon_d$ and half bandwidth, 
$w_d$, namely as
\begin{equation}
\label{eq:semic-d}
  D_d(\omega)  = \frac{2}{\pi w_d}\sqrt{1 - \frac{(\omega -\epsilon_d)^2}{w_d^2}}\:.
\end{equation}
In contrast the $s$-$p$ band is taken as having its center at zero and a large bandwidth, $w_s$,
\begin{equation}
\label{eq:semic-s}
  D_s(\omega)  = \frac{2}{\pi w_s}\sqrt{1 - \frac{\omega^2}{w_s^2}}\:.
\end{equation}
Thus, in addition to $\epsilon_a$ and the relevant hybridisation parameters the Anderson model is uniquely defined by the 
center and width of the $d$ band and by the width of the $s$-$p$ one. Furthermore, since we take the approximation that 
the $s$-$p$ band remains unchanged across the transition metal series its contribution to the integral of Eq.~(\ref{eq:be-2})
can be simply replaced by a constant, $E_{sp}$, specific for each adsorbate. 

In what follows the band parameters, $\epsilon_d$ and $w_d$, will be extracted from DFT calculations with a procedure
described in the next sections, while $\epsilon_a$ and $E_{sp}$ will be considered as fitting parameters. Finally, as far as
$V_{ad}$ is concerned, we will use a well-known strategy~\cite{Hammer2000} of considering the tabulated values extracted 
from the LMTO potential functions.~\cite{Andersen1985} These are essentially scaling rules, so that the hybridisation 
parameters are all known up-to a general scaling constant, $\beta$, which also will be fitted. 
{Note that the same scaling rules are also used for the NA model, which then requires the parameter $\beta$.}
{Note also that the band parameters, which in principle should be computed for each specific 
surface, are here replaced with those of the bulk compound. This is an approximation that allows us to perform a 
large-scale analysis of the entire transition metal space, but that makes our model insensitive to the surface details. 
The validity of such approximation will be discussed in section~\ref{sec:modfit}.}
\begin{table}[th]
  \centering
  {\renewcommand{\arraystretch}{1.5} % <---modify value for larger vspace
    \begin{tabular}{|l|l|l|l|l|l|l|l|}
      \hline
      Model & \multicolumn{4}{c|}{\color{Green} DFT} & \multicolumn{3}{c|}{\color{Red}FIT} \\
      \hline
      % Model & band center  & coupling  &       & Work function &  Impurity level & Constant energy & scaling\\
      NA & $\epsilon_d$ & $V_{ad}$  &       &        & $\epsilon_a$      & $E_{sp}$ & $\beta$\\
      \hline
      M1 & $\epsilon_d$ & $V_{ad}$  & $w_d$ &        & $\epsilon_a$      & $E_{sp}$ & $\beta$\\
      \hline
      M2 & $\epsilon_d$ & $V_{ad}$  & $w_d$ & $WF$ & $\epsilon_a+WF$ & $E_{sp}$ & $\beta$\\
      \hline
    \end{tabular}
  }
  \caption{The different models investigated in this work. Quantities indicated with `DFT' are directly extracted from DFT 
  calculations for bulk materials, while those in the `FIT' column are taken as fitting parameters. Note that none of the models
  is defined by more than three fitting parameters, specific for each adsorbate.}
  \label{tab:models}
\end{table}

A summary of the models investigated is presented in Table~\ref{tab:models}, where we separate the quantities that we will 
extract from DFT (`DFT' column) from those used as fitting parameters (`FIT' column). 
{NA is the original Newns-Anderson model, Eq.~(\ref{eq:be-na}), with $\epsilon_d$ taken as the $d$-band
center. In contrast, M1 and M2 are just numerically defined and essentially implement Eq.~(\ref{eq:be-2}). In M1 the adsorbate 
energy level is constant for each adsorbate, while in M2 its position is shifted by the experimental work-function of the metal, 
$WF$ (either experimental or extracted from DFT). Note that all the models require only three fitting parameters.}

\subsection{Density functional theory calculations}
\label{sec:DFT}

DFT calculations are performed for the 4$d$ transition metal series (Y to Cd), which is taken as benchmark
for our models and as training dataset for their fit. We use the all-electron FHI-AIMS code\cite{Blum2009}, since
its local-orbital basis set makes it more numerically efficient than plane-wave schemes in computing surfaces. A revised 
version of the Perdew-Burke-Ernzerhof (RPBE) exchange-correlation functional~\cite{PhysRevB.59.7413}, extensively 
tested for adsorption energies, is considered throughout this work, together with the `light' basis-set FHI-AIMS default. 
Tests for the more accurate `tight' basis set have revealed that the average error is minimal compared with that of the 
descriptor. 
 
For all the 4$d$ elemental phases we perform two sets of calculations either considering the experimental lattice structure
(see table \ref{tab:4d-1}) or a hypothetical {\it fcc} structure at the RPBE energy minimum. In both cases we construct 
4-to-6-layer thick slabs with surfaces along the [100], [110] and [111] directions. The lateral dimensions of the supercell is 
such that the surface contains a minimum of 4 atoms, so to minimise the interactions between the periodic images (single 
impurity limit). The reciprocal space is sampled with a 12$\times$12$\times$1 grid and the relaxation is converged when 
the forces are smaller than $5.0\times10^{-3}$~eV/\AA.

The binding energy is then calculated as
\begin{equation}
E_\mathrm{b} = E_\mathrm{ads+slab} - E_\mathrm{slab} - E_\mathrm{ads}
\end{equation}
where, $E_\mathrm{ads}$ is the DFT energy of the adsorbate alone (oxygen or sulfur)
{in its atomic form}, 
$E_\mathrm{slab}$ that of 
the `relaxed' slab, and $E_\mathrm{ads+slab}$ is the energy of the relaxed slab including the adsorbate at its equilibrium
position. We always relax the top layer of the slab when calculating either $E_\mathrm{slab}$ or $E_\mathrm{ads+slab}$. 
In both cases the lower layers are kept fixed to reduce the computational overhead. The orbital-resolved DOS for the 
bulk structure and for the various surfaces are always calculated to be used for fitting the models. In that case the 
Brillouin zone is sampled with a 144$\times$144$\times$144 and a 144$\times$144$\times$1 grid, respectively.

\begin{table}
\caption{\label{tab:4d-1} Summary table of the elementary phases of the 4$d$ transition-metal series investigated in
this work. For each element we report the atomic number, $Z$, the atomic configuration, the lattice structure of the
thermodynamically stable phase at room temperature, the most stable surface (the one investigated here), the experimental
work function, $WF$ (in eV), the Pauling electronegativity, $EN$, (for O and S this is 3.44 and 2.58 respectively), the
computed energy position of the $d$ band (in eV), $\epsilon_d$, the computed width of the $d$ band, $w_d$ (in eV).
Note that $\epsilon_{d}$ is taken with respect to the Fermi level, which is set to zero.}
\begin{tabular}{lllccclll} \hline\hline
  El. & $Z$ & Conf.           & Lattice & Surface    & $WF$ & $EN$ & $\epsilon_\mathrm{d}$ & $w_\mathrm{d}$  \\ \hline
  Y   & 39  & 5$s^2$4$d^1$    & hcp     & (0001) & 3.1                 & 1.22 & ~1.77 & 1.88 \\
  Zr  & 40  & 5$s^2$4$d^2$    & hcp     & (0001) & 4.05                & 1.33 & ~1.12 & 2.14 \\
  Nb  & 41  & 5$s^2$4$d^3$    & bcc     & (100)  & 3.95-4.87           & 1.60 & ~0.31 & 2.12 \\
  Mo  & 42  & 5$s^2$4$d^4$    & bcc     & (100)  & 4.36-4.95           & 2.16 & -0.18 & 2.19 \\
  Tc  & 43  & 5$s^2$4$d^5$    & hcp     & (0001) & 5.01                & 1.9  & -0.87 & 2.34 \\
  Ru  & 44  & 5$s^2$4$d^6$    & hcp     & (0001) & 4.71                & 2.2  & -1.76 & 2.14 \\
  Rh  & 45  & 5$s^2$4$d^7$    & fcc     & (100)  & 4.98                & 2.28 & -1.98 & 1.76 \\
  Pd  & 46  & 5$s^2$4$d^8$    & fcc     & (100)  & 5.22-5.6            & 2.2  & -1.96 & 1.41 \\
  Ag  & 47  & 5$s^2$4$d^9$    & fcc     & (100)  & 4.64                & 1.93 & -4.3 & 0.90 \\
  Cd  & 48  & 5$s^2$4$d^{10}$ & hcp     & (0001) & 4.08                & 1.69 & -8.95 & 0.45 \\
  \hline\hline
\end{tabular}
\end{table}

\subsection{Data extraction from AFLOWLIB.org}
\label{sec:aflow}

As explained before we have carried out novel RPBE-DFT calculations only for the 4$d$ transition-metal 
series, which has served as dataset for fitting the model. Once the model has been determined this has 
been run over an existing extremely large dataset of electronic structure calculations. In particular we 
have extracted data from the AFLOWLIB.org library~\cite{Curtarolo2012}. At present, this contains basic 
electronic structure information (DOS, band structure, etc.) for about 3.2 millions compounds, including
about 1,600 binary systems ($\sim$350,000 binary entries) and 30,000 ternary ones (2,400,000 ternary
entries). These have all been computed at the PBE level with the DFT numerical implementation coded in
the VASP package~\cite{PhysRevB.54.11169}, and with extremely standardised convergence criteria.
In particular, a subset of the AFLOWLIB.org data is for experimentally known compounds, namely for 
entries of ICSD.\cite{Zagorac2019} Our work investigates that particular subset.

{It must be noted that there may be an apparent inconsistency in constructing the models 
by using RPBE electronic structures and applying them to PBE data. However, one has to consider that
RPBE is a variation of PBE designed to improve over the energetics of chemisorption processes. The
two functionals remain relatively similar, and most importantly here, they produce rather close Kohn-Sham
spectra, and hence DOS. The variations in DOS between RPBE and PBE have very little influence on
the determination of the binding energy from our models, and certainly they generate errors much smaller 
than that introduced by not considering structural information in our descriptors (see next section).}

The AFLOWLIB.org library is accessible through a web-portal for interactive use, but most importantly 
through a RESTful Application Program Interface (API).\cite{TAYLOR2014178} This implements a query 
language with a syntax, where comma separated `keywords' (the material properties available) are followed 
by a `regular expression' to restrict the range or choice of the keywords. For instance the string
$\textrm{``Egap(1*,*1.6),species(Al,O),catalog(icsd)''}$ will give us a list of compounds containing Al and 
O and included in ICSD, whose band gap ranges in the interval [1eV, 1.6eV]. Such queries are submitted
to the server \url{http://aflowlib.duke.edu/search/API/?}. Here we have used the AFLOWLIB.org 
RESTful API, together with Python scripting, to search and extract the material information of 
transition-metal (i) elemental phases, and both (ii) binary and (iii) ternary alloys. In all cases we have
limited the search to metals only.

We have found that, in general, for a given material prototype AFLOWLIB.org contains multiple entries. 
Some correspond to different stable polymorphs, but also there is redundancy for a given lattice, where 
multiple entries differ by small variations of the lattice constants. These are typically associated
to independent crystallographic characterisations of the same material, taken under slightly different 
experimental conditions (temperature, pressure, etc.) or at different moments in time. Such small variations 
typically change little the electronic structure, so that for our purpose they provide no extra information. We 
have then removed such `duplicates' by grouping the compounds by lattice symmetry and total DFT energy. 
Then, for a given crystal structure we have selected the entry presenting the lowest energy. Such procedure
has returned us 88 elemental phases, 646 binary and about 50 ternary metallic alloys. For these we
have extracted the orbital-resolved DOS, which was then fitted to the semi-circular DOS of 
Eq.~(\ref{eq:semic-d}).

\begin{figure}[htb] %[!htb] %fig-2
\begin{center}
  \includegraphics[width=8cm]{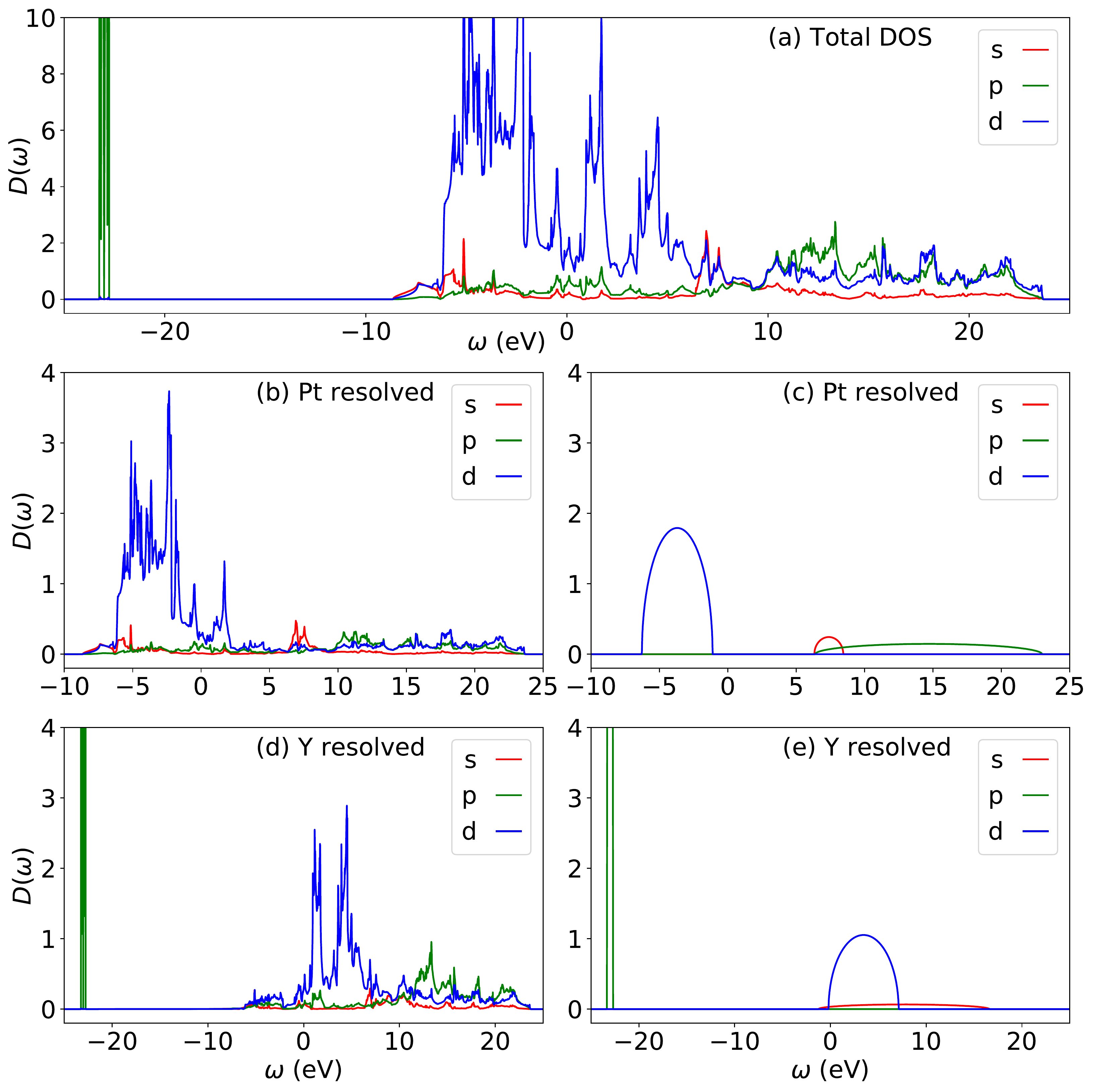}
\end{center}
\caption{(Color online): Orbital-resolved DOS for Pt$_2$Y$_1$ (ICSD 649861), as extracted from AFLOWLIB.org and its corresponding fits. 
In panel (a) we show the DFT-computed total orbital-resolved DOS. Panels (b) and (d), respectively are the element-resolved, orbital-resolved 
DOS for Pt and Y. In panels (c) and (e) we show our semi-circular fit to the Eq.~(\ref{eq:semic-d}) semicircular DOS.
}
\label{fig:dosfit}
\end{figure}

The fit proceeds by minimising the mean squared variance between the actual DFT-calculated 
DOS, $D_{\text{DFT}}(\omega)$, and the semicircular expression, $D_d(\omega, \epsilon_i, w_i)$, namely by 
minimising the following quantity
\begin{equation}
\Delta D=\left(D_{\text{DFT}}(\omega) - \sum_{i=1}^{n}\eta_i D_d(\omega, \epsilon_i, w_i)\right)^2\:.
\end{equation}
Here, $n$ is the number of semicircles used in the fit, while $\eta_i$, $\epsilon_i$, $w_i$ are the weights, centres and 
half widths of each semicircle, respectively. 
{We have initially used a variable number of semicircles to fit the DOS, but found that a single one
for each atomic orbital was always providing the best result. The fit extends to all species present in a compound and all 
orbital channels ($s$, $p$, $d$ and sometime $f$), but only data related to the $d$ band, $\epsilon_d$ and $w_d$, 
of all the species are retained when using the model.} An example of such fit is provided in Fig.~\ref{fig:dosfit} for 
Pt$_2$Y$_1$. 

\section{Results and Discussion}
\label{sec:result}
\subsection{Fitting the models}
\label{sec:modfit}
\begin{figure}[htb] %[!htb] %fig-2
\begin{center}
\includegraphics[width=8.5cm]{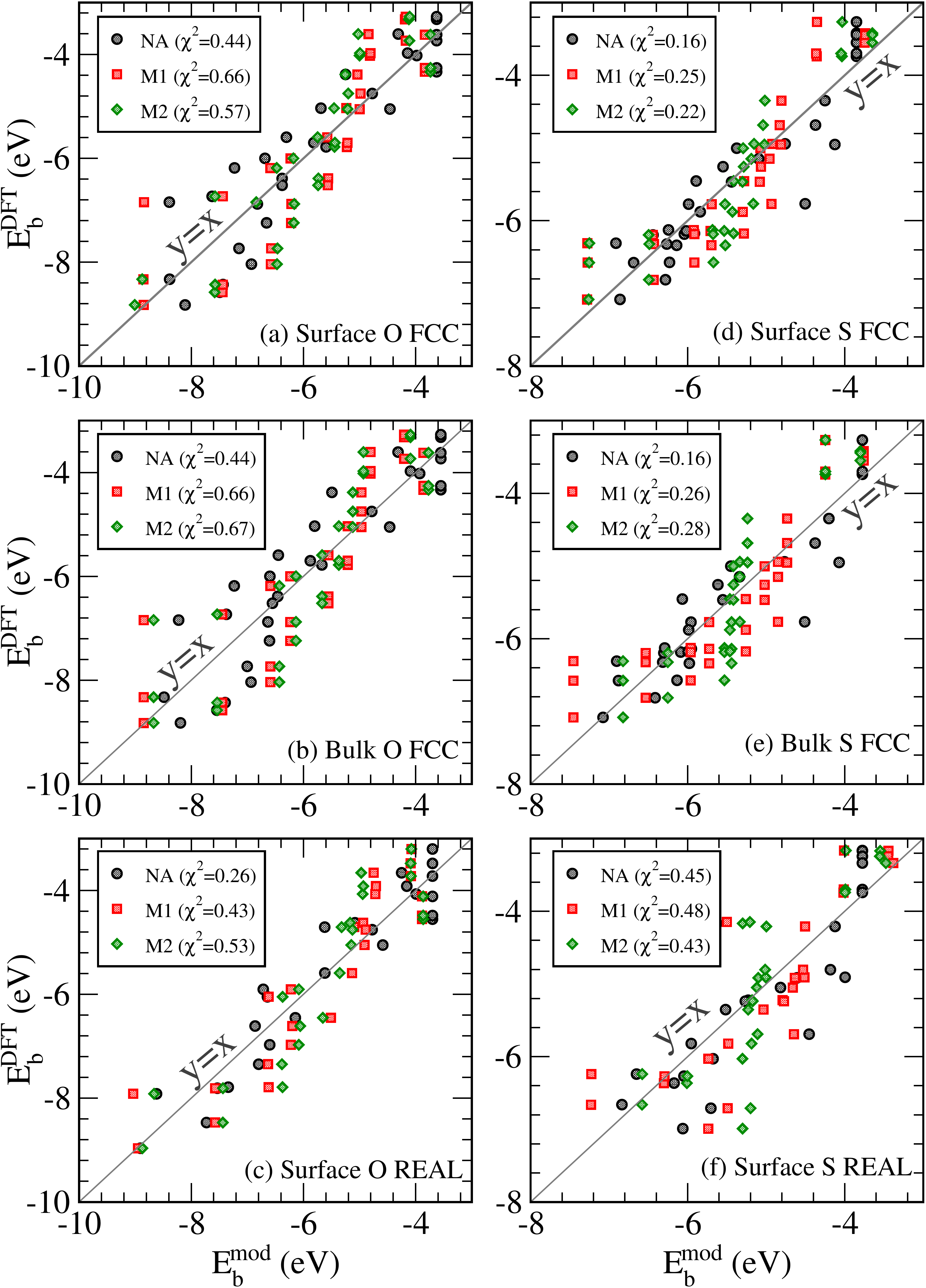}
\end{center}
\begin{center}
  \begin{tabular}{|l|l|l|l|l|l|l|}
    \hline
    model & $\sigma$: (a) & $\sigma$: (d) & $\sigma$: (b) & $\sigma$: (e) & $\sigma$: (c) & $\sigma$: (f) \\
          & O    & S    & O    & S    & O    & S \\
    \hline
    NA    & 0.66 & 0.40 & 0.66 & 0.40 & 0.51 & 0.67\\
    M1    & 0.81 & 0.50 & 0.81 & 0.51 & 0.66 & 0.70\\
    M2    & 0.76 & 0.47 & 0.82 & 0.53 & 0.73 & 0.66\\
    \hline
  \end{tabular}
\end{center}

\caption{(Color online) Comparison between the DFT binding energies and those estimated with the best fit
of the various models. The left column [panels (a)-(c)] is for oxygen, while the right column [panel (d)-(f)] for 
sulfur. Circles, squares and diamonds correspond to the NA, M1 and M2 model, respectively. The relative 
$\chi^2$ minimum of each model is shown in the legends (see text for details).
The table contains the binding-energy mean deviation, $\sigma = \sqrt{\chi^2(\epsilon_a, \beta, E_{sp})}$.
{The two top panels are for the {\it fcc} structures with the fit done over the DOS of the 
specific surfaces; the two middle are again for the {\it fcc} structures, but now we use the bulk DOS for the
fit; the two lower panels are for the actual experimental crystal structures of the compounds.}}
\label{fig:comp-model}
\end{figure}

Each of the three models introduced in the previous sections requires to determine three parameters, $\epsilon_a$, $\beta$ 
and $E_{sp}$, specific to each adsorbate. These are obtained by fitting the RPBE data for the 4$d$ transition-metal series. 
In particular we minimise the sum of the mean squared difference between the DFT binding energies, 
$E^\mathrm{DFT}_\mathrm{b}$, and those computed by the models, $E^\mathrm{mod}_\mathrm{b}(\epsilon_a, \beta, E_s)$, namely
\begin{equation}
\chi^2(\epsilon_a, \beta, E_{sp}) = \frac{1}{N_{s}}\sum_{i\in [\mathrm{Y-Cd}]} 
[(E^\mathrm{DFT}_\mathrm{b})_i - E^\mathrm{mod}_\mathrm{b}(\epsilon_a, \beta, E_{sp})_i]^2\;.
\end{equation}
where $N_s$ is the total number of surfaces taken over the [Y-Cd] range. For {\it fcc} bulk and surface, $N_s=33$, while for real structures 
we had $N_s=28$. The fits for O and S are performed independently, with $\epsilon_d$ taken from the RPBE calculations, $V_{ad}$ from the
scaling laws of Ref.~[\onlinecite{Andersen1985}] and $WF$ from experiments. Finally, the electronic filling of the $d$ band, $f$,
or equivalently the position of the Fermi energy are also taken from the RPBE calculations.

Our best fits are presented in Fig.~\ref{fig:comp-model}, where we show the model-predicted energies against the RPBE values
for both oxygen [panels (a)-(c)] and sulfur [panels (d)-(f)]. The table below the figure reports the mean absolute deviation of the
binding energy, $\sigma = \sqrt{\chi^2(\epsilon_a, \beta, E_{sp})}$. As we go down the rows in the figure, we have three different sets of 
fit, which differ for the choice of the DFT-calculated DOS taken to compute the $d$-band centre and bandwidth, and for the
target DFT energies. In the first row [panels (a) and (d)] the DOS is that of the surface atoms of the metals constrained to the {\it fcc}
lattice, and so are the target DFT energies. In the second-row panels [(b) and (e)] the DOS is that of the bulk {\it fcc} lattice, while
the target binding energies remain the same. Finally the last panels [(c) and (f)] use data for the metals in their thermodynamically 
stable structure (see Table~\ref{tab:4d-1}).

In general we find that all the models tend to perform better for oxygen than for sulphur, in particular when the actual equilibrium
structures are considered [panels (c) and (f)]. Note that the spread of the DFT RPBE binding energies for O is significantly larger 
than that of S (by about 2~eV), reflecting the same trend observed for the enthalpies of formation of oxides and sulphides (see 
Fig.~\ref{fig:1}). This means that a similar $\chi^2$ translates in a smaller relative error for O. Interestingly, while in the case of O
our best fit is obtained for the experimental structures, the opposite happens for S, for which the fit for the hypothetical {\it fcc} 
lattice is significantly more accurate. In fact, we find that the worst performance is obtained for S and the experimental structures, regardless
of the model used. This large error is associated to a significant scattering in the actual DFT data, in particular towards the beginning
of the series. For instance we find that when going from the most stable (0001) surface of {\it hcp} Y to the same for Zr the binding
energy marginally increases (becomes less negative), as expected from the larger occupation of the $d$ band. However, when moving 
to the most stable (100) surface of {\it bcc} Zr, $E_\mathrm{b}$ significantly decreases and in fact it becomes lower than that of
both Y and Zr. Clearly such behaviour cannot be captured by any of the models, since when going from Y to Zr to Nb the position of
$\epsilon_d$ monotonically increases (see Table~\ref{tab:4d-1}). Similar anomalies are found for Ru and Rh, although much
less pronounced. 

The much more pronounced spread in binding energies for sulphur can be attributed to its electronegativity, lower than that of
oxygen, and to the associated ability to form compounds involving transition metals over a broad range of stoichiometry. This
is particularly evident towards the beginning of the transition metal series. For instance, while Y forms only one stable oxide and
one sulphide, Y$_2$O$_3$ and Y$_2$S$_3$, so that it takes only the 3+ oxidation state, Zr has a single oxide, ZrO$_2$, but
can form sulphides with five different stoichiometries, ranging from Zr$_3$S$_2$ to ZrS$_3$ (see tables in Appendix \ref{materials_tables}). 
Most importantly the enthalpy of formation of these different sulphides varies significantly, from 0.77~/eV/atom for Zr$_3$S$_2$
to 1.99~eV/atom for ZrS$_2$. It is also interesting to note that even when there are stable oxides formed with the same transition
metal but with different stoichiometry, therefore yielding a different oxidation state for the metal, the fluctuation in enthalpy of 
formation remains small. For instance one can find NbO (Nb oxidation state 2+) with an enthalpy of formation of 2.17~eV/atom and 
Nb$_2$O$_5$ (Nb oxidation state 5+) with an enthalpy of formation of 2.81~eV/atom. 

A second important conclusion can be taken by looking at the first two rows of Fig.~\ref{fig:comp-model}, where the same DFT 
binding energies computed for the {\it fcc} lattice are modelled by using the band parameters of either the surface atoms [panels
(a) and (d)] or those of the bulk [panels (b) and (e)]. Clearly the two sets of fit present very similar errors, a fact that reflects the
small changes in band parameters when going from the surface to the bulk. Indeed such changes do exist and in fact there
is established evidence for binding energy shifts with $d$-band center shifts.~\cite{GREELEY2005104} However, while the inclusion
of these small corrections improve the fit when a relatively narrow range of metals is considered, they have little impact in our
case, that considers the entire transition metal series. Since our intention is to examine a very broad range of metallic alloys 
we can approximate the DOS of the surface with that of the bulk. This allows us to avoid performing surface calculations for the
several hundreds compounds previously selected. {An interesting possibility for improving on such assumption would 
be that of constructing a simple descriptor correlating the DOS narrowing at surfaces with the DOS of the bulk.}

{A more quantitative estimate of the accuracy of our model, at least for the elemental phases, can be
obtained by analysing in more detail the distribution of DFT binding energies for the 4$d$ transition-metal series across
the different surfaces of the actual structures and the hypothetical {\it fcc} ones. Such distribution is available in 
Fig.~\ref{fig:comp-model}, and it is re-plotted as a function of the atomic number in Fig.~\ref{fig:fig10} in 
appendix~\ref{DFTbindenergy}. From the figures we notice that the spread in values is of the order of 1~eV across the series, 
with the exception of Tc and both Mo and Nb, but only in the case of S. Clearly, Tc is not a matter of concern, since it is 
radioactive and forms a rather limited number of known binaries. Mo and Nb are more problematic and effectively set the 
accuracy of our model, which is of the order of $\pm$1~eV.}

Finally, we notice that when comparing the different models we find little difference in accuracy, with the original NA model
performing slightly better than both M1 and M2. {This fact is somehow counterintuitive, since one expects a
better fit for models including more parameters. We attribute such behaviour to the fact that here we apply the models to
a very broad distribution of binding energies, for which the fluctuations of the DFT values are relatively large. Over such range
the accuracy of the model is mainly driven by the $d$-band center, while finer details, such as the bandwidth, appear to have
little impact. Note that in literature there are several examples of model improvements associated to descriptors, which include
more information about the band shape~\cite{Xin2014}. These, however, are related to narrow subsets of compounds, for 
which the $d$ band center changes little, and the binding energy is driven by more subtle features of the electronic
structure. For this reason in the remaining of the paper we will consider the NA model only.}

\subsection{Binding energies of binary alloys to O and S}
\label{sec:be-bin-o}

\begin{figure}[t] %[!htb] %fig-2
\begin{center}
\includegraphics[width=0.48\textwidth]{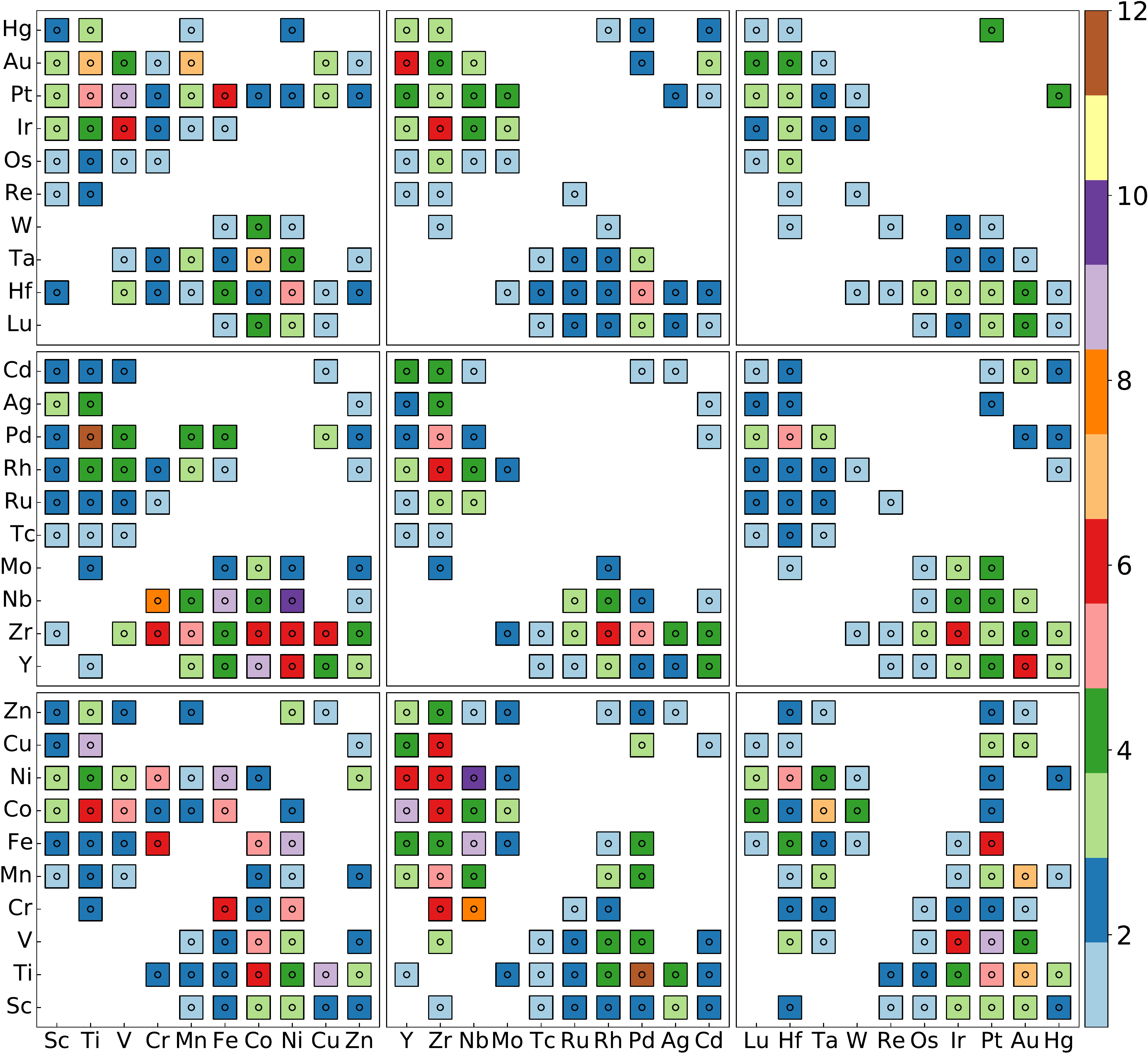}
\end{center}
\caption{(Colour online): Distribution of stable phases across the entire transition-metal binary system map.
The number of compounds for a given binary system is colour coded.}
\label{fig:ba-bin}
\end{figure}

We now discuss the trends in reactivity of transition-metal binary alloys to oxygen and sulfur. Out of the 30 transition metals there 
are $30\times29 / 2 = 435$ binary systems, a number that needs to be compared with the 646 binary intermetallic 
compounds found at the interception between the AFLOWLIB.org and the ICSD databases. A more detailed view of the 
chemical distribution of such 646 compounds can be obtained by looking at Fig.~\ref{fig:ba-bin}, where we graphically plot 
the number of stable phases for each of the 435 binary systems. Firstly, we note that there are several binary systems for 
which no single compound is found. This does not necessarily means that the two elements are not miscible, but simply that 
there is no stable ordered crystalline phase, for which a full crystallographic characterisation is available. This is, for instance,
the case of the Hf-Zr system; the two elements are miscible at any concentration, but the thermodynamically stable phase
is a solid-state solution across the entire composition diagram. A similar situation is found for many binary systems made of 
elements belonging to the same group or to adjacent groups, namely along the right-going diagonal of the matrix of Fig.~\ref{fig:ba-bin}.
In contrast, there is a much stronger tendency to form stable intermetallic phases in binary systems comprising
an early ($d^{0}-d^{3}$) and a late ($d^{7}-d^{10}$) transition metal. For instance Ti-Pd is the system presenting 
the largest number, namely 12 of stable phases.

Next we move to discuss the trend in binding energies to oxygen and sulfur across the binary-system space.
Clearly the binding energy is an object that depends on both the chemical composition and the stoichiometry of
a compound, namely for a binary alloy it is a four dimensional function. Thus for the $A_xB_y$ binary one
has $E_\mathrm{b}(A_xB_y) = f(A, x, B, y)$. We then proceed in the following way. For each binary system
$A$-$B$ we analyse all existing stoichiometry, $A_xB_y$, and compute all the possible binding energies by
running the NA descriptor against the partial DOS of all inequivalent bulk atomic sites. Then, we plot on
a matrix analogous to that of Fig.~\ref{fig:ba-bin} the minimum, $E_\mathrm{b}^\mathrm{min}$, 
and maximum, $E_\mathrm{b}^\mathrm{max}$, binding energy found for that system, namely
\begin{figure*}[t] %[!htb] %fig-2
  \begin{center}
    \includegraphics[width=0.48\textwidth]{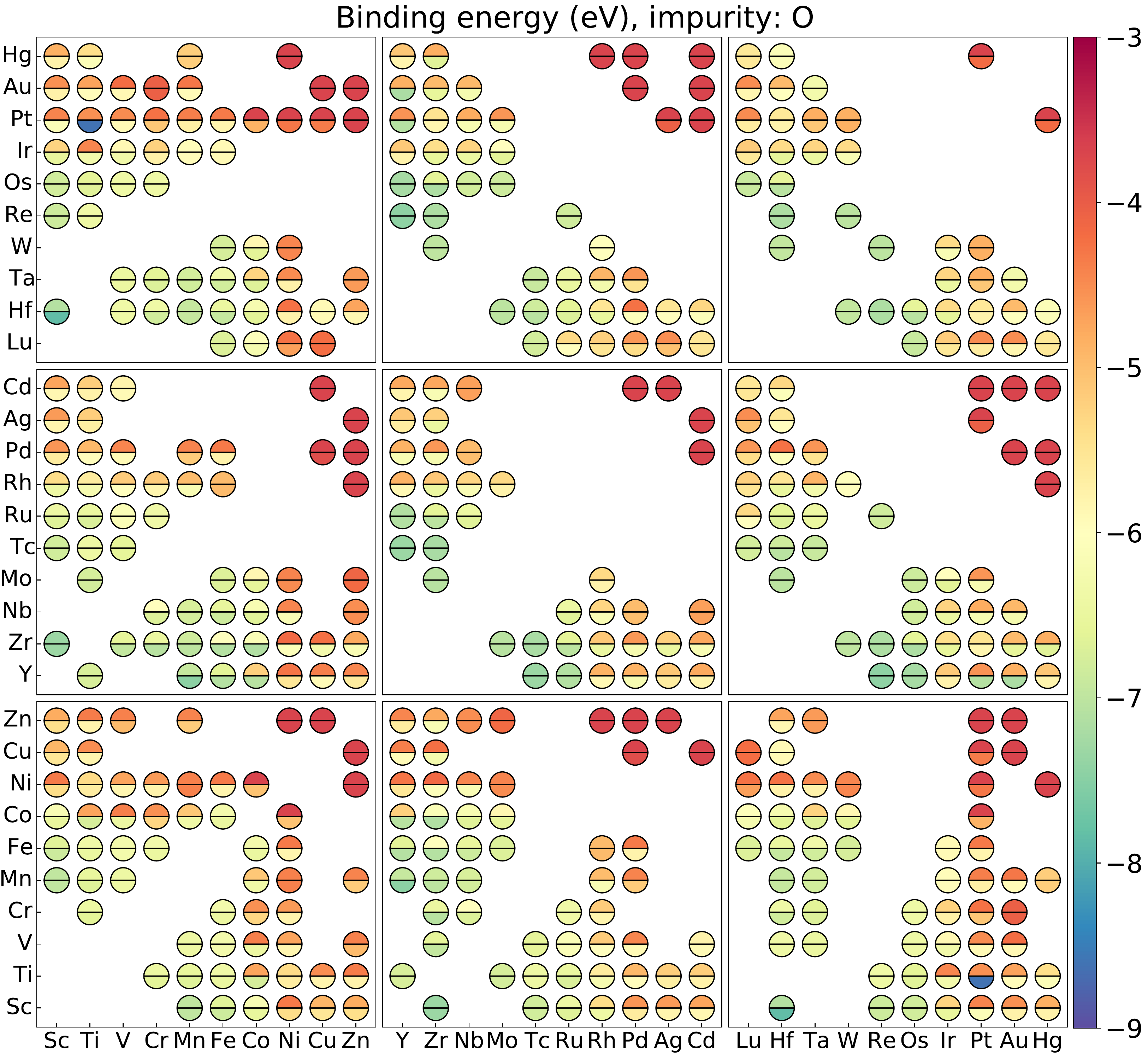}
    \includegraphics[width=0.48\textwidth]{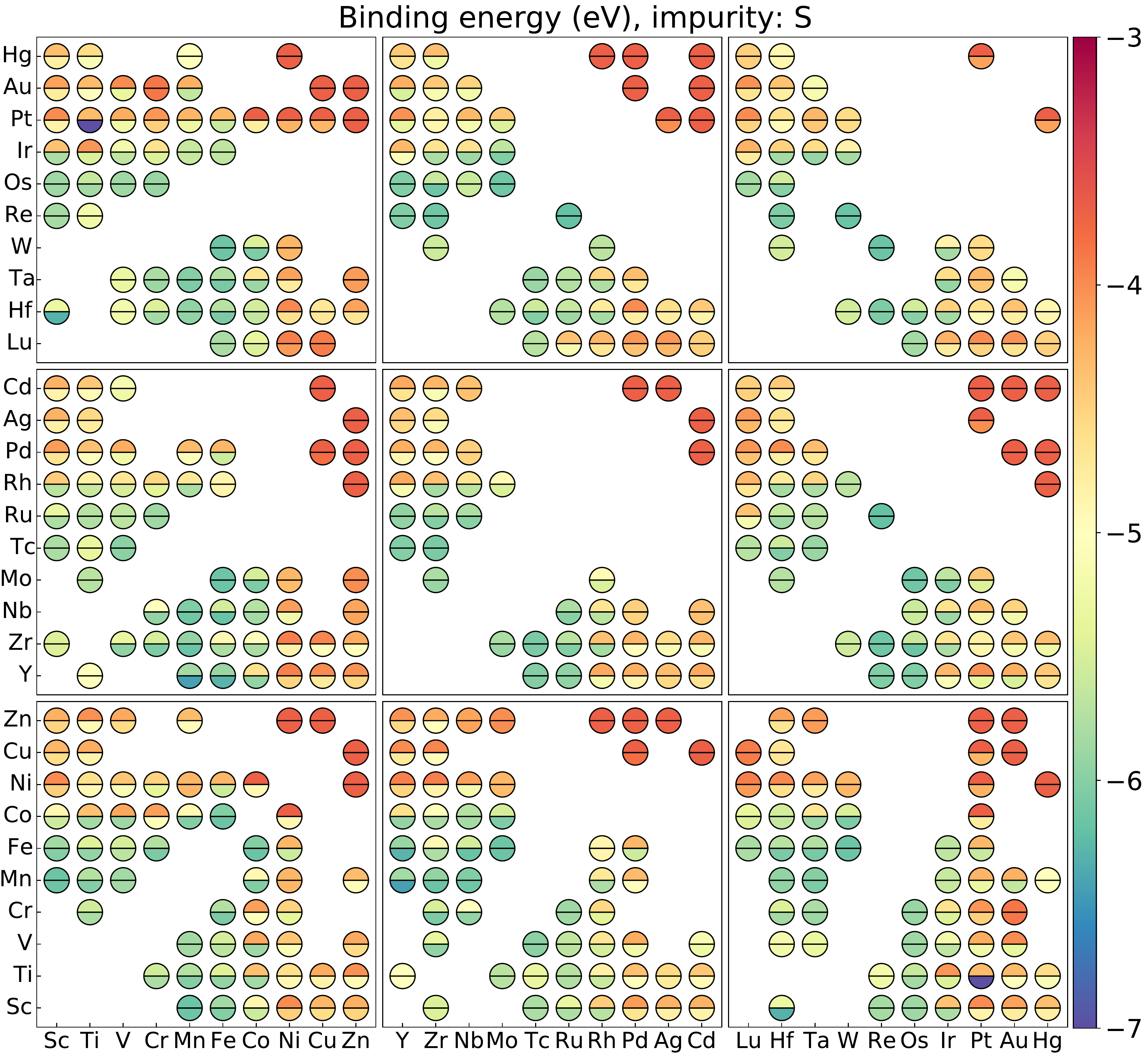}
  \end{center}
  \caption{(Colour online) Binding-energy map across the transition-metal binary systems.
    The left panel is for oxygen, and right panel for sulfur.
    For each given binary the colour of the two semicircles encodes the value of $E_\mathrm{b}^\mathrm{min}$ 
    and $E_\mathrm{b}^\mathrm{max}$ (energies are in eV according to the heat-map scale depicted on the side
    of each picture). These are taken across the different stoichiometry and possible binding sites for any given 
    binary system.}
  \label{fig:be-bin-os}
\end{figure*}
\begin{align}\label{Emax}
  E_\mathrm{b}^\mathrm{max}(A:B) &= \max_{x,y}\{E_\mathrm{b}(A_xB_y)\}\:,\\ 
    E_\mathrm{b}^\mathrm{min}(A:B) &= \min_{x,y}\{E_\mathrm{b}(A_xB_y)\}\:.
\end{align}

The results of this analysis for both oxygen (left-hand side graph) and sulphur (right-hand side graph) are presented in 
Fig.~\ref{fig:be-bin-os}, where $E_\mathrm{b}^\mathrm{min}$ and $E_\mathrm{b}^\mathrm{max}$ occupy the two halves 
of a circle and are encoded as a heat map. In the figure red tones indicate a weak binding energy, thus low reactivity, while 
the green/blue ones are for strong binding and high reactivity. When the two halves of a particular circle appear approximately
of the same colour there is little variance in binding energy across stoichiometry and binding sites, while a strong contrast means 
that for that binary system there are extremely reactive sites (for some stoichiometry) together with weak ones (for the same 
compound or for different ones). Note that Fig.~\ref{fig:be-bin-os} just depicts the trend in reactivity, but it cannot be taken as an 
absolute measure of the binding energy across the binary space. In fact, it is not guaranteed that either $E_\mathrm{b}^\mathrm{min}$ 
or $E_\mathrm{b}^\mathrm{max}$ are actually accessible. For instance one may have the situation in which the most reactive site 
of a given binary system is associated to an unstable surface, so that it will be hardly available in practice. 

Nevertheless, Fig.~\ref{fig:be-bin-os} provides valuable insights into the reactivity to oxygen and sulfur of the transition-metal alloys.
As expected one finds the less reactive compounds among the alloys made of late ($d^{7}-d^{10}$) transition metals, 
regardless of the period they belong to. For these the binding energy is small and so is the variance across stoichiometry
and binding sites. The binding energy then becomes increasingly more negative as we move along the diagonal of
the plots towards early ($d^{0}-d^{3}$) transition metals. For this class the variance still remains relatively low owning
to the fact that the position of the center of the $d$ band is similar for elements in near groups. A different situation
is encountered when one moves off the plot diagonal, namely towards alloys combining an early and a late transition 
metal. In this case we typically find both strongly and weakly reactive sites. An extreme example is found for the Pt-Ti 
system, where an $E_\mathrm{b}^\mathrm{min}$ of -8.62~eV is found for the Ti site of Ti$_3$Pt$_1$, while a 
$E_\mathrm{b}^\mathrm{max}$ of -4.53~eV corresponds to Pt in Ti$_1$Pt$_8$. This last group is of particular interest, 
in particular if one can find binary alloys with an overall weak reactivity, since the constituent elements do not include only
expensive noble metals. It is also important to note that there is correlation in the variance between $E_\mathrm{b}^\mathrm{min}$ 
and $E_\mathrm{b}^\mathrm{max}$ and the abundance of compounds in a particular binary system (see Fig.~\ref{fig:ba-bin}). 
In fact, we find that more abundant binary systems, which have potentially a larger stoichiometric space (and thus a larger variance
in possible inequivalent binding sites), have a larger colour contrast in Fig.~\ref{fig:be-bin-os}. 

Let us now compare the two panels of Fig.~\ref{fig:be-bin-os} for oxygen and sulphur. This is an important exercise, since often in
an ambient-condition corrosion process O and S compete for the same binding site, thus that their relative reactivities may determine 
the final products of reaction and the overall reaction rate. In general, a bird-eye view of the data suggests rather similar qualitative 
chemical trends for sulphur and oxygen. However, a closer looks reveals a few differences: (i) the overall binding energies for sulphur 
are lower than those of oxygen (note that the scales in the two panels of Fig.~\ref{fig:be-bin-os} are different); (ii) the spread, or the 
variation over the minima and maxima (basically the colour contrast between the upper and lower semi-circles), for sulphur is typically 
larger  than for oxygen (e.g. in Hg:Pt). 

\subsection{Reactivity of binary alloys to elemental O and S}
\label{sec:be-bin-o}

We are now going to develop a simple criterion for estimating, on a more qualitative ground, the relative reactivity of a given binary
system to S and O. The idea is to compare the predicted reaction rates for O and S absorption and to evaluate these from our
computed binding energies. For simplicity here we take O$_2$ and S$_2$ as the main reactants, so that the reaction of interest
is: $TM+1/2\mathrm{O}_2\rightarrow TM_\mathrm{O}$, where $TM$ indicate the transition metal and $TM_\mathrm{O}$ the 
transition metal with one O adsorbed (the same holds for S). The enthalpy of reaction, $\delta E^n$ ($n=$O, S), can then be simply 
written as $\delta E^n=E_\mathrm{b}^n+1/2E_\mathrm{atom}^{n_2}$, where $E_\mathrm{atom}^{n_2}$ is the experimental atomization enthalpy for either O$_2$ (5.1~eV) or S$_2$ (4.4~eV), { and $E_\mathrm{b}^n$ is the binding 
energy of the $n$ specie}. 
Importantly, the enthalpy of reaction is often found to be linearly correlated to the activation
energy, at least in the case of small molecules interacting with late transition metals. These so-called Br\o{}nsted-Evans-Polanyi 
relations~\cite{BEP1,BEP2} thus establish a simple connection between a thermodynamical quantity, the enthalpy of reaction, and 
a dynamical one, the activation energy, $E_\mathrm{act}^n$. Thus, one has $E_\mathrm{act}^n=\gamma \delta E^n+\xi$, where the
coefficients $\gamma$ and $\xi$ are, in principle, specific of any given reactant. Finally, the reaction rate, $\kappa$, is solely determined 
by the activation energy via the usual Arrhenius expression, $\kappa=\nu \mathrm{e}^{-E_\mathrm{act}/k_\mathrm{B}T}$, where $\nu$ 
is the frequency factor, $T$ the temperature and $k_\mathrm{B}$ the Boltzmann constant.

The crucial point in the discussion is the observation that the scaling coefficient entering the Brønsted-Evans-Polanyi relations,
$\gamma$ and $\xi$, are universal for different classes of molecules and/or bonds~\cite{Norskov2002,Michelides2003}. For instance for 
simple diatomic homonuclear molecules (e.g. O$_2$, N$_2$) one finds $\gamma\sim0.95$ and $\xi\sim2$~eV. By assuming that
the same relation is valid also for S$_2$, we can then write an expression for the ratio between the reaction rates of O and S,
namely
\begin{equation}\label{rates}
\frac{\kappa_\mathrm{S}}{\kappa_\mathrm{O}}=\frac{\nu_\mathrm{S}}{\nu_\mathrm{O}}\:\mathrm{exp}\left({-\frac{\gamma(\delta E^\mathrm{S}-\delta E^\mathrm{O})}{k_\mathrm{B}T}}\right)\:.
\end{equation}
If one wants to use Eq.~(\ref{rates}) to determine the relative reaction rate at ambient conditions, then we will write $T=300$~K,
and a further simplification can be made by assuming that the frequency factors for O$_2$ and S$_2$ are similar,
$\nu_\mathrm{S}/\nu_\mathrm{O}\sim1$. Finally, considering that the typical S concentration in the lower atmosphere is about 1~ppm, 
one expects similar corrosion to S and O when their reaction rates are in the ratio $\kappa_\mathrm{S}/\kappa_\mathrm{O}\sim10^6$. 
This leads to the condition
\begin{equation}\label{rates2}
1=\frac{1}{2}\:\mathrm{exp}\left({-\frac{\gamma(E^\mathrm{S}_\mathrm{b}-E^\mathrm{O}_\mathrm{b})}{0.025~\mathrm{eV}}}\right)\:.
\end{equation}
We can than conclude that a given transition metal alloy will corrode equally to S and O when $E^\mathrm{S}_\mathrm{b}\sim E^\mathrm{O}_\mathrm{b}$, 
otherwise the reactivity will be dominated by oxidation. 

\begin{figure}[t] 
\begin{center}
\includegraphics[width=0.48\textwidth]{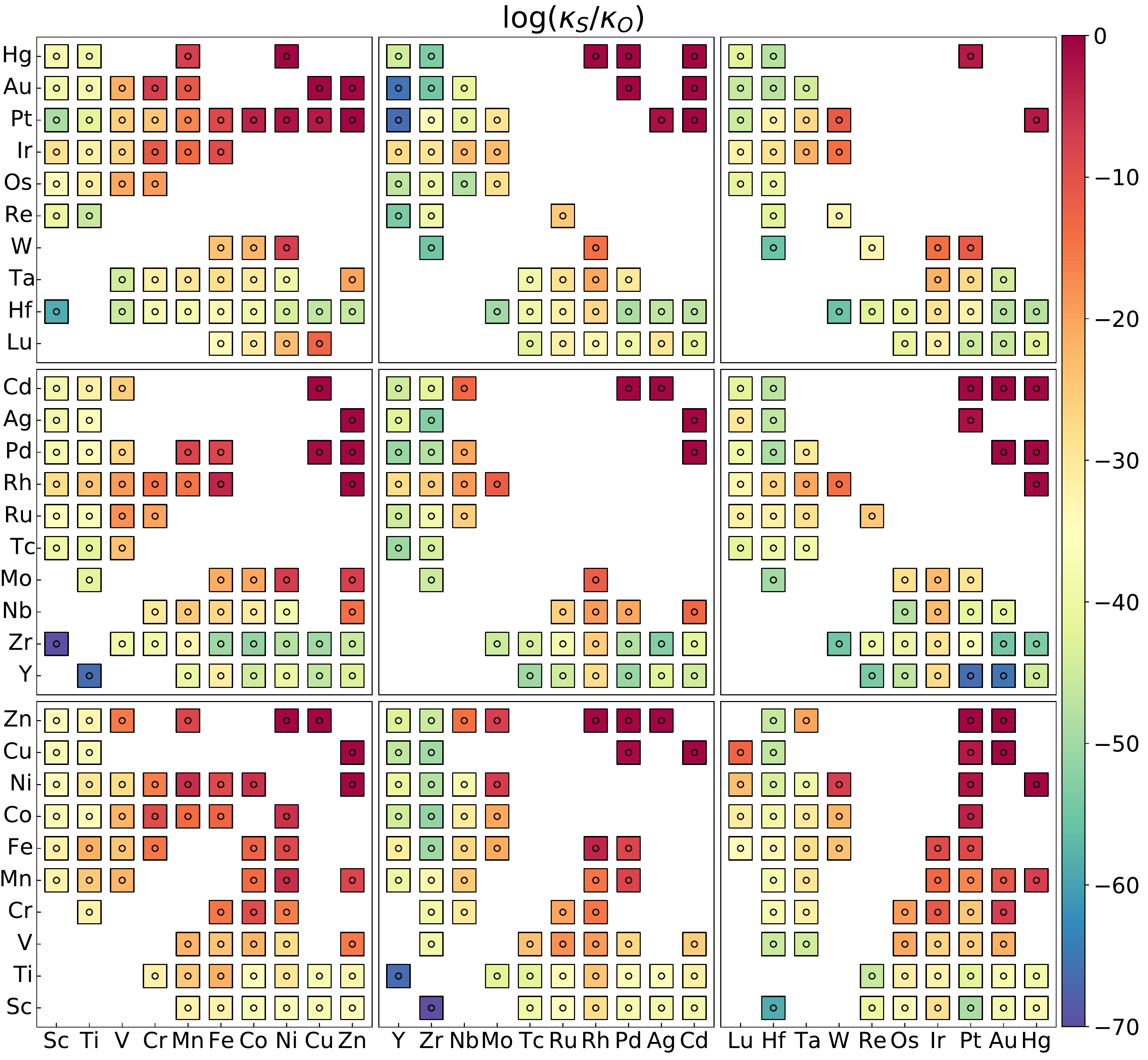}
\end{center}
\caption{(Colour online): Reaction rates ratio $\kappa_\mathrm{S}/\kappa_\mathrm{O}$, Eq.~(\ref{rates}), for the binary systems investigated. 
Here $\ln(\kappa_\mathrm{S}/\kappa_\mathrm{O})$ is plotted as a heat map for all the binary compositions by taking as binding energy 
$E^\mathrm{max}_\mathrm{b}$.}
\label{RateRatio}
\end{figure}

It is important to note, however, that in the atmosphere S is present mainly in the SO$_2$ and H$_2$S form, and not as S$_2$. Unfortunately 
Brønsted-Evans-Polanyi relations are currently unavailable for SO$_2$ and H$_2$S so that a more quantitative analysis of the ambient relative 
reactivity of O and S cannot be carried out. Nevertheless, the ratio $\kappa_\mathrm{S}/\kappa_\mathrm{O}$ of Eq.~(\ref{rates}) can serve as a 
useful descriptor to analyse the relative reactivity to S and O of a binary system. This analysis is carried out in Fig.~\ref{RateRatio}, where we plot 
$\ln(\kappa_\mathrm{S}/\kappa_\mathrm{O})$ over our binary space, and we take $E^\mathrm{max}_\mathrm{b}$ as binding energy.

As expected the $\ln(\kappa_\mathrm{S}/\kappa_\mathrm{O})$ map closely resembles that of the binding energies (see Fig.~\ref{fig:be-bin-os}),
with lower $\kappa_\mathrm{S}/\kappa_\mathrm{O}$ ratios for the late transition metals (particularly in the 5$d$ period), while the reactivity
to O is always largely dominant. From the figure it is clear that the condition $\kappa_\mathrm{S}/\kappa_\mathrm{O}>10^{-6}$, which 
makes the ambient corrosion to S stronger than that to O, is met only for a rather limited number of binary systems. In fact, this seems to be
unique to alloys with both atomic species having more than 10 valence electrons (Ni, Pd and Pt). However, it is important to note that this
analysis is based on $E^\mathrm{max}_\mathrm{b}$, so that it is not specific of a particular compound, but simply explore trends in the binary
space. One can than still have binary compounds, where the electronic interplay between the atomic species results in lower binding energies, 
and therefore reactivity, with respect to the binary system they belong to. This analysis is carried out next.

\subsection{Reactivity gain for binary compounds}
\label{sec:gain}

Our task here is to identify those binary compounds, in which the binding energy of the different inequivalent sites differs the most from
that of the corresponding single-element phases. In other words we wish to find those binary alloys for which the bond formation between
chemically different atoms alters the most the position of the $d$ band with respect to that of the elemental phase. For a generic $A_xB_y$ 
binary alloy such property can be captured by the ``elemental energy shift'', a descriptor defined as
\begin{equation}
\Delta E_\mathrm{el}=E^n_\mathrm{b}-E^n_\mathrm{el}\:,
\end{equation}
where $E^n_\mathrm{b}$ is the binding energy for the species $n$ ($n=A, B$) in the binary alloy and $E^n_\mathrm{el}$ is the binding
energy of the elemental phase of $n$. In particular, for any binary compound we compute the maximum and minimum value of $\Delta E_\mathrm{el}$.
Here $\Delta E_\mathrm{el}^\mathrm{max}$ (typically a positive value) corresponds to the particular site whose binding energy has been
increased the most in forming the alloy (the site is less reactive than the same element in its elemental phase), while $\Delta E_\mathrm{el}^\mathrm{min}$ 
(typically a negative value) is for the site whose binding energy has been reduced the most (the site is more reactive than in its elemental phase).
These two quantities are listed in Table~\ref{Tabgain} for the 24 compounds presenting the largest $\Delta E_\mathrm{el}^\mathrm{max}$ and
$\Delta E_\mathrm{el}^\mathrm{min}$ for both O and S. In the same table we also list the composition-averaged binding energy, $\tilde{E}_\mathrm{b}$, 
defined as $\tilde{E}_\mathrm{b}=w_xE^A_\mathrm{b}+w_yE^B_\mathrm{b}$, where $w_x=x/(x+y)$ [$w_y=y/(x+y)$]. This latter energy provides a rough 
estimate of the global reactivity of a particular compound.
\begin{table}[th]
\caption{\label{Tabgain} List of the binary compounds possessing binding sites, whose binding energy present large deviations with respect to that of
their corresponding elemental phases. All energies are in eV. In the brackets we report the particular element associated to $\Delta E_\mathrm{el}$.}
  \centering
  \resizebox{0.495\textwidth}{!}{
    \begin{tabular}{c|c|c|c|c|c|c}
      \hline \hline
      \toprule
      {} & \multicolumn{3}{c|}{ {\bf Oxygen}} & \multicolumn{3}{c}{ {\bf Sulphur}} \\ \hline
      {\bf Comp} & $\tilde{E}_\mathrm{b}$ & $\Delta E_\mathrm{el}^\mathrm{max}$ & $\Delta E_\mathrm{el}^\mathrm{min}$ & $\tilde{E}_\mathrm{b}$ & $\Delta E_\mathrm{el}^\mathrm{max}$ & $\Delta E_\mathrm{el}^\mathrm{min}$ \\\hline
      \midrule
              RhY & -5.87 &     4.44 (Y) &   -3.18 (Rh) & -4.63 &     2.01 (Y) &   -0.83 (Rh) \\
        Cu$_{5}$Y & -4.38 &     4.44 (Y) &   -4.08 (Cu) & -3.99 &     2.01 (Y) &   -1.76 (Cu) \\
        Ir$_{2}$Y & -5.14 &     4.43 (Y) &   -2.75 (Ir) & -4.32 &     2.00 (Y) &   -0.44 (Ir) \\
             AuSc & -5.74 &    3.86 (Sc) &   -4.08 (Au) & -4.75 &    1.96 (Sc) &   -2.10 (Au) \\
       Ni$_{5}$Sc & -4.31 &    3.86 (Sc) &   -3.63 (Ni) & -3.99 &    1.96 (Sc) &   -1.70 (Ni) \\
             AuLu & -5.82 &    3.84 (Lu) &   -4.25 (Au) & -4.64 &    1.60 (Lu) &   -1.88 (Au) \\
       Au$_{3}$Lu & -4.72 &    3.84 (Lu) &   -4.08 (Au) & -4.13 &    1.60 (Lu) &   -1.71 (Au) \\
        Co$_{3}$Y & -5.22 &     3.67 (Y) &   -3.40 (Co) & -4.61 &    -0.37 (Y) &    0.44 (Co) \\
             PtZr & -5.82 &    3.58 (Zr) &   -3.58 (Pt) & -4.80 &    1.74 (Zr) &   -1.61 (Pt) \\
       Ni$_{3}$Zr & -4.43 &    3.58 (Zr) &   -2.85 (Ni) & -4.04 &    1.74 (Zr) &   -1.29 (Ni) \\
              IrW & -5.36 &     3.24 (W) &   -1.76 (Ir) & -4.83 &     2.02 (W) &   -0.87 (Ir) \\
       AgHf$_{2}$ & -5.99 &    3.22 (Hf) &   -3.44 (Ag) & -4.79 &    1.41 (Hf) &   -1.63 (Ag) \\
       Au$_{3}$Hf & -4.96 &    3.22 (Hf) &   -5.06 (Au) & -4.37 &    1.41 (Hf) &   -2.70 (Au) \\
        Ni$_{4}$W & -4.44 &     3.16 (W) &   -3.33 (Ni) & -4.27 &     1.95 (W) &   -2.50 (Ni) \\
             IrNb & -5.26 &    3.08 (Nb) &   -1.57 (Ir) & -4.63 &    1.59 (Nb) &   -0.47 (Ir) \\
       Cd$_{3}$Nb & -4.69 &    3.08 (Nb) &   -3.94 (Cd) & -4.36 &    1.59 (Nb) &   -2.63 (Cd) \\
        Ir$_{3}$Y & -5.79 &     3.03 (Y) &   -2.55 (Ir) & -5.08 &     0.74 (Y) &   -0.32 (Ir) \\
             PdTa & -5.47 &    3.02 (Ta) &   -3.43 (Pd) & -4.69 &    1.47 (Ta) &   -1.88 (Pd) \\
       Pd$_{3}$Ta & -4.59 &    3.02 (Ta) &   -3.44 (Pd) & -4.33 &    1.47 (Ta) &   -2.42 (Pd) \\
       Pt$_{3}$Ta & -4.81 &    3.02 (Ta) &   -3.76 (Pt) & -4.28 &    1.47 (Ta) &   -1.71 (Pt) \\
       Rh$_{2}$Ta & -4.89 &    3.02 (Ta) &   -2.39 (Rh) & -4.52 &    1.47 (Ta) &   -1.41 (Rh) \\
             AuMn & -5.19 &    2.99 (Mn) &   -2.97 (Au) & -5.00 &    2.59 (Mn) &   -2.59 (Au) \\
       Ir$_{2}$Lu & -5.55 &    2.95 (Lu) &   -2.23 (Ir) & -4.74 &    0.80 (Lu) &   -0.12 (Ir) \\
 Cu$_{3}$Ti$_{2}$ & -5.04 &    2.89 (Ti) &   -3.34 (Cu) & -4.54 &    1.46 (Ti) &   -2.10 (Cu) \\
      \bottomrule
      \hline\hline
    \end{tabular}
  }
\end{table}

From the table we find, as somehow expected, that compounds formed from elements placed at the different edges of the $d$-metal
period present the largest $\Delta E_\mathrm{el}$. In general one finds that the binding energy of the most reactive elements, 
typically Y, Sc, Lu, Zr, Hf, W, Nb and Ta, is drastically reduced (up to 4~eV) with respect to that of the corresponding elemental phase.
At the same time, $E^n_\mathrm{b}$ of the least reactive element increases, often by a relatively similar amount. Such variations 
are significantly more pronounced when considering binding to O than to S, mostly because the binding energies to O are larger
and because their dependence on the $d$-band filling factor is more pronounced (see Fig.~\ref{fig:1}). 
{Interestingly, we can identify compounds whose composition-averaged binding energy is relatively low, 
$\tilde{E}_\mathrm{b}>-4.5$~eV for O, and similar for O and S (within some fraction of eV), and at the same time present inequivalent 
sites that bind drastically differently from their element phases. These are mostly Ni-based intermetallics such as 
Ni$_5$Sc ($\tilde{E}_\mathrm{b}^\mathrm{O}=$-4.31~eV, $\tilde{E}_\mathrm{b}^\mathrm{S}-\tilde{E}_\mathrm{b}^\mathrm{O}=$0.32~eV), 
Ni$_4$W (-4.44~eV, 0.17~eV), 
Ni$_3$Zr (-4.43~eV, 0.39~eV), 
and also 
Pd$_3$Ta (-4.59~eV, 0.36~eV) and
Cu$_{5}$Y (-4.38~eV, 0.39~eV).}

\subsection{Ternary alloys}
\label{sec:be-tern-os}
Finally we turn our attention to the ternary compounds. In this case the set available is significantly smaller than what found for the
binaries, and in fact the same search criterion used before now returns us only 50 ternary phases. This may look a bit surprising, 
since ICSD approximately counts about 40,000 binary and 75,000 ternary phases~\cite{Zagorac2019}. However, here we are considering 
only compounds made of transition metals. These are then prone to form solid state solutions or highly disordered phases~\cite{Curtarolo2019}, 
whose structures are typically not part of ICSD. Furthermore, we have only included the compounds that are both in 
ICSD and AFLOWLIB.org library~\cite{Curtarolo2012}, namely at the intersection of the `real (ICSD)' (the subset of ICSD reporting 
experimentally determined structures) and the `{\it ab initio} (AFLOWlib)' database. In any case, the ternaries considered can be 
found from the union of the sets S$_{3d} = $ [Sc-Zn], S$_{4d} = $~[Y-Zr, Mo, Ru, Pd-Cd] and S$_{5d} = $ [Lu-Hf, W-Re, Pt-Hg],
namely they may contain any of the 3$d$ element and a selection of the 4$d$ and 5$d$, with a preference for either early or late 
transition metals. 

\begin{figure}[htb] %[!htb] %fig-2
\begin{center}
  \includegraphics[width=\columnwidth]{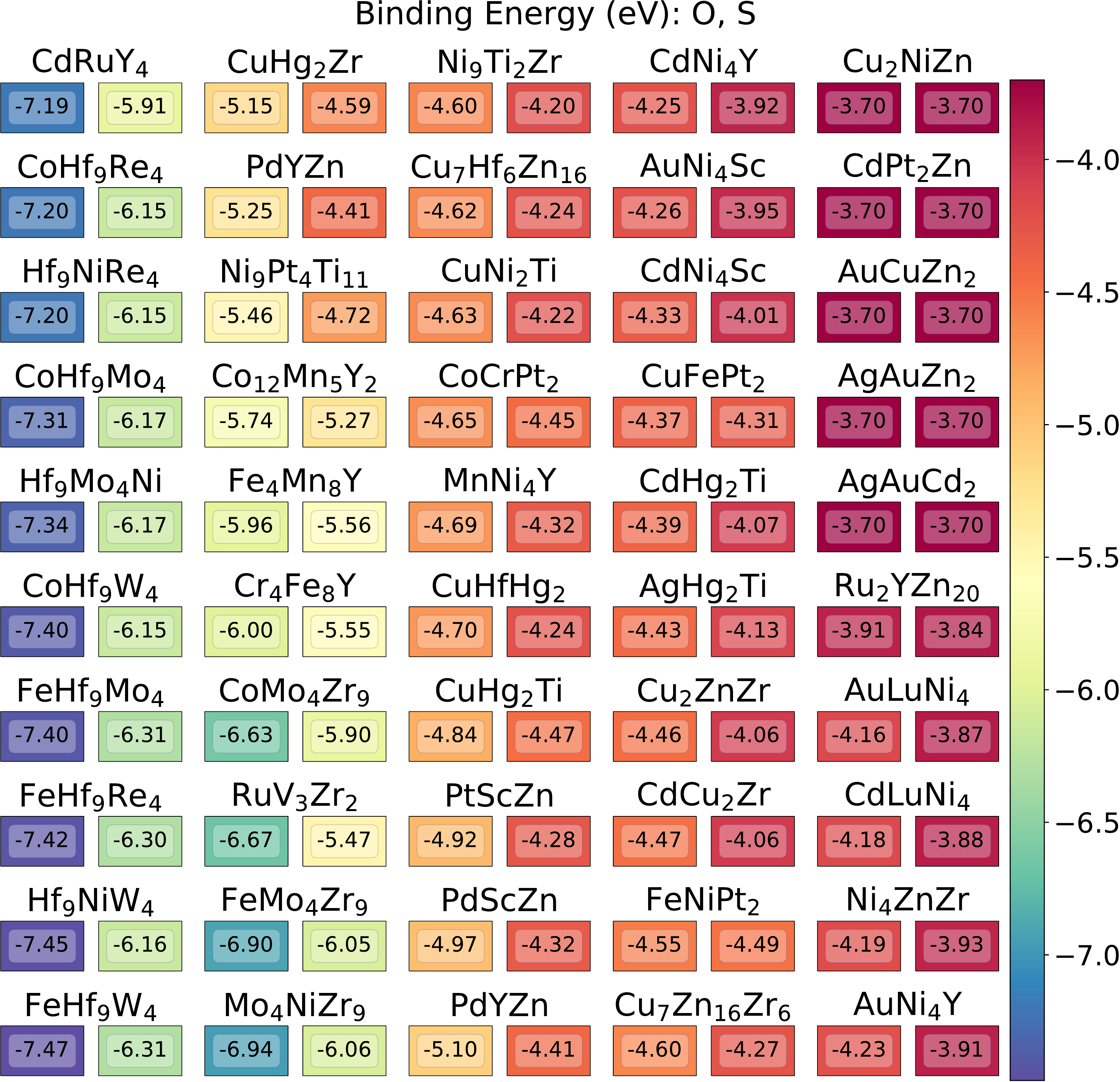}
\end{center}
\caption{(Colour online): Composition-averaged binding energy for all the ternary compounds investigated.
The two rectangles below each compound refer to O (left-hand side) and S (right-hand side), respectively,
and include the composition-averaged binding energy. The boxes are then colour coded with the same quantity
fo an easy visualization.}
\label{fig:be-tern}
\end{figure}

In Fig.~\ref{fig:be-tern} we show the list of these ternary compounds sorted by their composition-averaged binding energy, while 
details of the site-dependent binding energy together with the associated elemental energy shifts are provided in Table~\ref{tab:tern-os}
in the appendix. In general, as expected, the ternary phases showing shallower $\tilde{E}_\mathrm{b}$ are those including late 
transition metals, often going beyond the noble ones (e.g. Zn and Cd). More interestingly, the subset for which $\tilde{E}_\mathrm{b}$
is approximately the same for O and S are those with an average electronic configuration close to $s^2d^9$, namely that of 
Cu, Ag and Au. These, for instance include, Cu$_2$NiZn, CdPt$_2$Zn, AuCuZn$_2$ and AuCuCd$_2$. Among them, Cu$_2$NiZn 
appears particularly interesting, since it mimics the electronic structure of a noble metal, without including expensive elements. In 
contrast, at the opposite side of the distribution we find alloys with a dominant early transition-metal composition, for which the binding 
energy is deep and asymmetric between O and S. 

\subsection{Model validation for binary and ternary alloys}
\label{sec:be-tern-os}
{In closing this results section we are finally coming back to the question of the accuracy of our model
and the limits of its predictions. We have already remarked (see section~\ref{sec:modfit}) that the 
spread in the DFT binding energies across different surfaces for elemental 4$d$ transition-metal compounds
is in the region of 2~eV ($\pm$1~eV). Here we aim at validating such error for binary and ternary alloys.}

{To this goal we have selected four binary compounds and one ternary and computed the DFT binding energies
for O and S for several different surfaces. The compounds in questions are AgZr (ICSD number 605996), 
AgZr$_3$ (58392), CuPd (181913), Cu$_3$Pd (103084) and CuHfHg$_2$ (102969). In particular we have
selected two phases from the Ag-Zr binary system, as elemental Ag and Zr provide strongly and weakly coupling 
binding sites, respectively; and two compounds from the Cu-Pd system, since it is a low binding-energy one, and hence
interesting for applications. Finally, we have considered CuHfHg$_2$, since it includes elements with a 
broad range of binding strengths to O and S. For those we have computed the binding energies at the (100),
(110) and (111) surfaces, and whenever inequivalent, at the (001) one. Note that some of these compounds present
a layered structure, so that different surface terminations are possible. In this case we have computed the binding 
energy for all the inequivalent terminations. The calculations then proceed as for the elemental phases, by finding the 
minimum energy binding site, and its corresponding, $E_\mathrm{b}$.} 
\begin{figure}[t] %[!htb] %fig-2
\begin{center}
\includegraphics[width=0.45\textwidth]{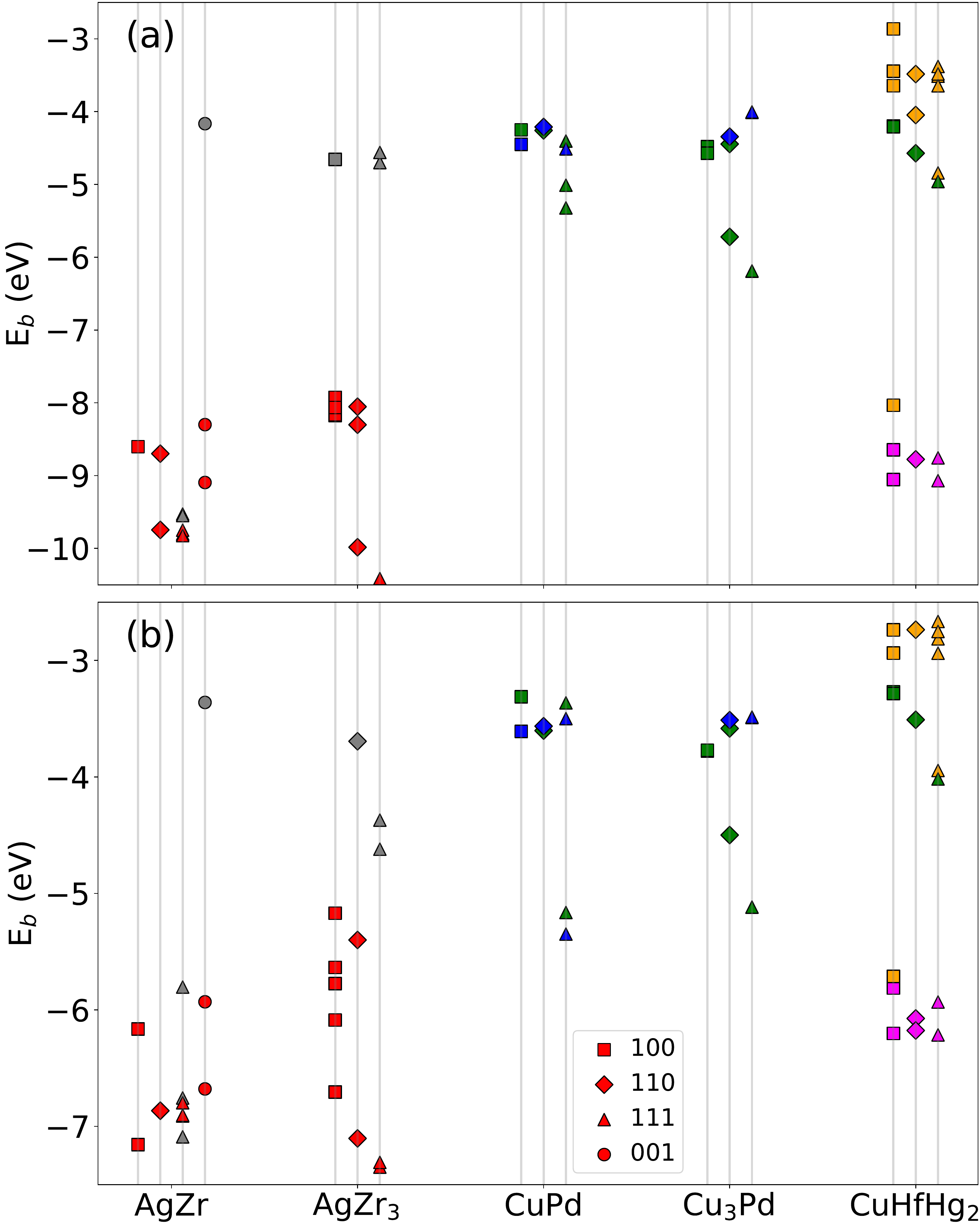}
\end{center}
\caption{(Colour online): DFT-computed binding energies of O (a) and S (b) for a number 
of selected binary and ternary compounds:  AgZr (ICSD \#605996) and AgZr$_3$ (ICSD \#58392); CuPd (ICSD \#181913), 
Cu$_3$Pd (ICSD \#103084) and CuHfHg$_2$ (ICSD \#102969). For each compound we report the binding energy 
of different surfaces and different absorption sites. Note that for the same surface orientation there may be different 
possible terminations. Different symbols correspond to different surfaces (the colour is that of the dominant binding 
specie). The colour code is as following: Ag = grey, Cu = green, Hf = magenta, Hg = orange, Pd = blue, Zr = red.}
\label{fig:fig9}
\end{figure}

{Our results are summarised in Fig.~\ref{fig:fig9}, where we report all the computed binding energies
and we colour-code the specific binding sites (the dominant site in the case the adsorbant coordinates with atoms 
belonging to different species). Let us start from the Ag-Zr system. In general this presents a bimodal distribution, with 
Zr-dominated binding sites showing low binding energies and Ag-dominated ones binding in a much weaker way.
In particular, the Zr sites have binding energies ranging from -10~eV to -8~eV for O and from -7~eV to -5~eV for S,
while the Ag ones are around -4.5~eV for O and -3.5~eV for S. These values have to be compared with those
predicted by our model by using the bulk DOS (see Fig.~\ref{fig:be-bin-os}). For the Zr-Ag system the model returns
a binding energy range of [-7, -5]~eV for O and [-6.5,-4.5]~eV for S. Thus, we find that our model is capable of
predicting the low binding-energy side of our distribution relatively well, while it appears to miss some Zr binding
sites with extremely low binding energy (it predicts the upper side of the Zr distribution). These are associated with
electron-depleted Zr-dominated surfaces, namely having electronic structure rather different from that of the bulk. }

{The situation is significantly better for the Cu-Pd system, where DFT returns a binding energy range
of [-6,-4]~eV for O and [-5,-3.5]eV for S, to be compared with a rather uniform model prediction of -4~eV for both
O and S. Such result is not surprising, since the electronic structure of Pd and Cu are rather similar, so that
large fluctuations at surfaces are not expected. Finally, when looking at our ternary compound we find a 
tri-modal distribution of binding energies with those associated to Hf at -9~eV (-6~eV), those to Cu at -4.5~eV
(-3.5~eV) and those to Hg at -3.5~eV (-3~eV) for O (S). For ternary alloys our model was used only to predict the
composition-averaged binding energy, which is -4.7~eV for O and -4.24~eV for S. The composition-averaged
$E_\mathrm{b}$ computed from the DFT values are -5.13~eV for O and -3.88~eV for S, namely in quite good 
agreement with the model.}

\section{Conclusion and outlook}
We have here investigated the propensity to oxidation and tarnishing of a large set of binary and ternary intermetallic
compounds. These were selected among existing phases, as reported in the ICSD and AFLOWLIB.org databases. Such
large-scale screening was enable by taking the binding energy to O and S as a proxy to the first stage of the corrosion 
process and by the definition of a descriptor. More specifically, we have utilised the well-known $d$-band-position concept and
constructed a descriptor with the associated parameters being fitted to DFT calculations for the 4$d$ transition metal series. 
A number of variants of the original Newns-Anderson model were evaluated before choosing one based on an analytical 
semi-elliptical density of states. Such descriptor was then put to work against the electronic structures contained in 
AFLOWLIB.org, after appropriate fitting, to investigate trends in the binding energy across the compositional space. 

In general, we have found the binding to O to be significantly stronger than that to S, a fact that follows closely the behaviour
of the enthalpy of formation of oxides and sulphides. Such difference, however, gets reduced as one moves across the
transition metal period towards the $s^2d^9$ atomic configuration ($s^1d^{10}$ in the solid state), characteristic of Cu,
Ag and Au. A somewhat similar situation is found for binary and ternary compounds. In this case, however, the presence of
chemically and structurally inequivalent sites complicate the analysis, which is better performed by computing for each 
compound and binary system the largest and smallest binding energies. This reveals binary systems presenting the 
co-existence of strongly and weakly binding sites, composed of an early and a late transition metal. At the same time,
binary phases made of elements belonging to adjacent groups display little variation in binding energy with the particular
binding site. 

The thermodynamical information contained in the binding energy can be then converted into a proxy for the reactivity by using
the so-called Br\o{}nsted-Evans-Polanyi relations. These are directly available for O$_2$ and have been extrapolated also
to S$_2$. By using such relations we have established that, at ambient conditions (temperature and relative S/O abundance), 
early-stage corrosion to O and S compete only when the binding energies are comparable, otherwise oxidation always 
appears to dominate. This first situation takes place only for late, and usually expensive, transition-metal alloys. However, in
the ternary space this seems to be possible also for a handful of alloys presenting an average $s^1d^{10}$ configuration,
but not necessarily including Ag and Au, such as Cu$_2$NiZn.

Overall our work provides a first rough navigation map across the binary and ternary transition metal composition space,
which is useful to categorise materials according to their propensity to oxidation and tarnishing. 
{Certainly, our method has several limitations, which need to be overcome in order to establish a high-throughput
quantitative theory of surface reactivity across such vast chemical and structural space. Firstly, we need to improve over our
binding-energy prediction ability. For instance, our descriptor is completely agnostic to the specific surface and absorption site. 
One first improvement may be that of running the NA model over the specific surface DOS, an operation that, however, will require
DFT surface calculations for the entire database, a numerically daunting task. In that case one may include additional features of the 
surface DOS into the model, which is likely to become more accurate.~\cite{Xin2014} An alternative strategy is to develop models 
taking into account the possibly bonding geometry of the different bonding sites. If machine-learning schemes~\cite{Wang2020}
can be constructed, this may limit the number of surface calculations to perform.}

{Secondly, we need to establish a more solid link between the binding energy and the surface chemical activity. 
In this case, one has to validate a new set of Br\o{}nsted-Evans-Polanyi relations for O- and S-containing atmospheric 
gases,~\cite{Aas2018} such as H$_2$S or SO$_2$. This will involve performing reaction path calculations over a range of
surfaces. The task is relatively straightforward for elemental phases, but becomes much more complex in the case of binary
and ternary alloys. Also in this case a machine-learning strategy generalising or replacing completely the Br\o{}nsted-Evans-Polanyi 
approach may be a possible solution. A few examples in such direction exist~\cite{Stocker2020,Lee2020}, but to date the field 
remains quite uncharted.}

\section*{ackowledgement}

We thank Corey Oses, Cormac Toher and Stefano Curtarolo for support with the AFLOWLIB.org API.
This work is supported by by Science Foundation Ireland (Amber center 12/RC/2278) and by Nokia Bell Lab. 
Computational resources have been provided by the supercomputer facilities at the Trinity Center for High 
Performance Computing (TCHPC) and at the Irish Center for High End Computing (ICHEC), projects 
(tcphy108c and tcphy120c). 

\appendix
\section{Semi-circular density of state}\label{appdx-1}

We consider the Anderson impurity model for the adsorbate problem, defined as follows.
We consider the impurity with onsite level, $\epsilon_a$, coupled to the $s$-$p$  and the $d$ 
band of a transition metal. The band dispersions for the metal are $\epsilon^s_{\bf k}$ and
$\epsilon^d_{\bf k}$. One can then write the following Hamiltonian [see Eq.(\ref{eq:Ham})], 
\begin{align}\nonumber
  \label{eq:Ham1}
  H &= \sum_{\sigma} \underbrace{\epsilon_a a^{\dagger}_{\sigma}a_{\sigma}}_{\text{Impurity}}
      + \sum_{{\bf k}\sigma}
      \underbrace{\epsilon^s_{\bf k}s^{\dagger}_{{\bf k}\sigma}s_{{\bf k}\sigma} + \epsilon^d_{\bf k}d^{\dagger}_{{\bf k}\sigma}d_{{\bf k}\sigma}}_{s-d~\text{band dispersion}}\\
    &+ \sum_{{\bf k}\sigma}
      \underbrace{V^s_{\bf k}a^{\dagger}_{\sigma}s_{{\bf k}\sigma} + V^d_{\bf k}a^{\dagger}_{\sigma}d_{{\bf k}\sigma} + \textrm{h.c.}}_{\text{Impurity-band coupling}}\:,
\end{align}
where $V^s_{\bf k}$ and $V^d_{\bf k}$ are the coupling integrals of the $s$-$p$ and $d$ band to the impurity level.
The operator $a^{\dagger}_{\sigma}$ ($a_{\sigma}$) creates (annihilate) an electron in the impurity level,
while the $s^{\dagger}_{{\bf k}\sigma}$ ($s_{{\bf k}\sigma}$) and $d^{\dagger}_{{\bf k}\sigma}$ ($d_{{\bf k}\sigma}$)
do the same for an electron in the bulk state $|{\bf k}\sigma\rangle _{s,d}$ of the $s$-$p$ and $d$ bands respectively. 
Since there is no mixing in the spins, for the moment we drop the spins label $\sigma$.

Now lets calculate the impurity density of state (DOS). We define the impurity and mixed Green's functions as follows,
\begin{align}
  \label{eq:gfdef}
  G_{aa}(t) &= -i\theta(t)\langle [a(t),a^{\dagger}(0)] \rangle\:,\\
  G_{ad}({\bf k},t) &= -i\theta(t)\langle [d_{\bf k}(t),a^{\dagger}(0)] \rangle\:,\\
  G_{as}({\bf k},t) &= -i\theta(t)\langle [s_{\bf k}(t),a^{\dagger}(0)] \rangle\:,
\end{align}
where, $d_{\bf k}(t) = e^{iHt}d_{\bf k}e^{-iHt}$ and so on. The equation of motion for these
Green's function are
\begin{align}
  i\frac{\partial G_{aa}(t)}{\partial t} &= \delta(t) + i\theta(t)\langle \left[[H,a(t)],a^{\dagger}(0)\right] \rangle\:,\\
  i\frac{\partial G_{ad}({\bf k},t)}{\partial t} &= i\theta(t)\langle \left[[H,d_{{\bf k}}(t)],a^{\dagger}(0)\right] \rangle\:,\\
  i\frac{\partial G_{as}({\bf k},t)}{\partial t} &= i\theta(t)\langle \left[[H,s_{{\bf k}}(t)],a^{\dagger}(0)\right] \rangle\:.
\end{align}
It is easy to see, that $[H,a] = -\epsilon_a a - \sum_{\bf k}V^s_{\bf k}s_{\bf k} - \sum_{\bf k}V^d_{\bf k}d_{\bf k}$,
$[H,s_{\bf k}] = -\epsilon^s_{\bf k}s_{\bf k} - V^{s*}_{\bf k}a$, and
$[H,d_{\bf k}] = -\epsilon^d_{\bf k}d_{\bf k} - V^{d*}_{\bf k}a$, where
$V^{s*}_{\bf k}$ is complex conjugate of $V^{s}_{\bf k}$. By using these identities, the equations of motion
simplify to,
\begin{align}\nonumber
  i\frac{\partial G_{aa}(t)}{\partial t} &= \delta(t) + \epsilon_a G_{aa}(t) \\
  &+ \sum_{\bf k}\left( V^d_{\bf k}G_{ad}({\bf k},t) + V^s_{\bf k}G_{as}({\bf k},t)\right)\:,\\
  i\frac{\partial G_{ad}({\bf k},t)}{\partial t} &= \epsilon^d_{\bf k}G_{ad}({\bf k},t) + V^{d*}_{\bf k}G_{aa}(t)\:,\\
  i\frac{\partial G_{as}({\bf k},t)}{\partial t} &= \epsilon^s_{\bf k}G_{as}({\bf k},t) + V^{s*}_{\bf k}G_{aa}(t)\:,
\end{align}
which in Fourier space become algebraic equations,
\begin{align}
  \label{eq:gimp}
  \nonumber
  (\omega - \epsilon_a)G_{aa}(\omega) &= 1 + \\
                                      & \sum_{\bf k}\left( V^d_{\bf k}G_{ad}({\bf k},\omega) + V^s_{\bf k}G_{as}({\bf k},\omega)\right)\:,
\end{align}

\begin{align}
  \label{eq:gmix}
  (\omega - \epsilon^d_{\bf k})G_{ad}({\bf k},\omega) &= V^{d*}_{\bf k}G_{aa}(\omega)\:,\\
  (\omega - \epsilon^s_{\bf k})G_{as}({\bf k},\omega) &= V^{s*}_{\bf k}G_{aa}(\omega)\:.
\end{align}

Now, by substituting $G_{as}({\bf k},\omega)$ and $G_{ad}({\bf k},\omega)$ from Eq.~(\ref{eq:gmix})
into Eq.~(\ref{eq:gimp}) we obtain the impurity Green's function
\begin{align}\nonumber
  (\omega - \epsilon_a)G_{aa}(\omega) &= 1 + \\
 & \sum_{\bf k}\left( \frac{|V^d_{\bf k}|^2}{\omega -\epsilon^d_{\bf k} + i\eta} + \frac{|V^s_{\bf k}|^2}{\omega -\epsilon^s_{\bf k} + i\eta} \right)G_{aa}(\omega)\:,
\end{align}
or, simplifying
 \begin{align}
   G_{aa}(\omega) &= {1 \over \omega -\epsilon_a - \sum_{\bf k}\left( \frac{|V^d_{\bf k}|^2}{\omega -\epsilon^d_{\bf k} + i\eta} + \frac{|V^s_{\bf k}|^2}{\omega -\epsilon^s_{\bf k} + i\eta} \right)}\\
   &= {1 \over \omega -\epsilon_a - \Sigma(\omega)}\:,
 \end{align}
where $\Sigma(\omega)$ is the self energy given by
 \begin{align}
   \Sigma(\omega) = \sum_{\bf k}\left( \frac{|V^d_{\bf k}|^2}{\omega -\epsilon^d_{\bf k} + i\eta} + \frac{|V^s_{\bf k}|^2}{\omega -\epsilon^s_{\bf k} + i\eta} \right)\:.
\end{align}
Now consider the imaginary part of the self energy ($\lim \eta \to 0$), which is readily related to the DOS,
\begin{align}
  \nonumber
  \text{Im}\Sigma(\omega) = -\sum_{\bf k}\left( \frac{|V^d_{\bf k}|^2\eta}{(\omega -\epsilon^d_{\bf k})^2 + \eta^2}
  + \frac{|V^s_{\bf k}|^2\eta}{(\omega -\epsilon^s_{\bf k})^2 + \eta^2} \right)\\
  = \pi\sum_{\bf k}\left( |V^d_{\bf k}|^2\delta(\omega - \epsilon^d_{\bf k}) + |V^s_{\bf k}|^2\delta(\omega - \epsilon^s_{\bf k}) \right)\:.
  \label{eq:sig-i}
\end{align}

If we assume the couplings to be independent of ${\bf k}$, namely $V^d_{\bf k}=V_d$ and $V^s_{\bf k}=V_s$,
we have the following
\begin{align}
  \nonumber
  \text{Im}\Sigma(\omega) &= \pi V_d^2\sum_{\bf k}\delta(\omega - \epsilon^d_{\bf k}) + \pi V_s^2\sum_{\bf k}\delta(\omega - \epsilon^s_{\bf k})\\
  &= \pi V_d^2 D_d(\omega) + \pi V_s^2 D_s(\omega) = \pi\Delta(\omega)\:,
\end{align}
where $D_s(\omega)$ and $D_d(\omega)$, are the density of states of the $s$-$p$ and $d$ bands. Thus, $\text{Im}\Sigma(\omega) = \pi\Delta(\omega)$, and using this, we have the real part of the self energy (say, $\Lambda(\omega)$) as
\begin{align}\nonumber
  \label{eq:sig-r}
  \text{Re}\Sigma(\omega) &= \Lambda(\omega) = P\int_{-\infty}^{\infty}{\Delta(\omega')d\omega' \over \omega - \omega'}\\
  &= V_d^2 P\int_{-\infty}^{\infty}{D_d(\omega')d\omega' \over \omega - \omega'} + V_s^2 P\int_{-\infty}^{\infty}{D_s(\omega')d\omega' \over \omega - \omega'}\:,
\end{align}
where $P$ denotes the principle part of the integral. Let us denote
\begin{align}
  \Lambda_d(\omega) = P\int_{-\infty}^{\infty}{D_d(\omega')d\omega' \over \omega - \omega'}\text{;~}
  \Lambda_s(\omega) = P\int_{-\infty}^{\infty}{D_s(\omega')d\omega' \over \omega - \omega'}\:,
\end{align}
so that we have $\Lambda(\omega) = V_d^2\Lambda_d(\omega) + V_s^2\Lambda_s(\omega)$. Thus we have
obtained the total self energy as a function of the $s$-$p$ and $d$ bands DOS
\begin{align}
  \Sigma(\omega) = \Lambda(\omega) - i\pi\Delta(\omega)\:.
\end{align}
The impurity Green's function can then be simplified to
\begin{align}
  G_{aa}(\omega) = {1 \over \omega -\epsilon_a - \Lambda(\omega) + i\pi\Delta(\omega)}\:.
\end{align}
Finally the impurity density of state, $D_a(\omega) = -\frac 1\pi\text{Im}G_{aa}(\omega)$, is
given by
\begin{align}
  D_a(\omega) = { \Delta(\omega) \over (\omega - \epsilon_a - \Lambda(\omega))^2 + \pi^2\Delta(\omega)^2 }\:.
\end{align}

If we choose a semi-circular DOS model, with center at $\epsilon_d$ and half bandwidth $w_d$
for the $d$ band, and center at 0 and half bandwidth $w_s$ for $s$ band, the two DOSs will write
\begin{align}
  D_d(\omega) = \frac{2}{\pi w_d}\sqrt{1 - \frac{(\omega -\epsilon_d)^2}{w_d^2}},\\
  D_s(\omega) = \frac{2}{\pi w_s}\sqrt{1 - \frac{\omega^2}{w_d^2}}\:.
\end{align}
Then an exact expression for $\Lambda_d(\omega)$ and $\Lambda_s(\omega)$ can be evaluated to
\begin{align}
  \Lambda_d(\omega + \epsilon_d) =
  \begin{cases}
    \frac{2}{w_d^2} (\omega + \sqrt{\omega^2-w_d^2}) & \text{$\omega < -w_d$}\\
    \frac{2}{w_d^2} \omega & \text{$|\omega|\le w_d$}\\
    \frac{2}{w_d^2} (\omega - \sqrt{\omega^2-w_d^2}) &\text{$\omega > w_d$}\\
  \end{cases}
\end{align}

\begin{align}
  \Lambda_s(\omega) =
  \begin{cases}
    \frac{2}{w_s^2} (\omega + \sqrt{\omega^2-w_s^2}) & \text{$\omega < -w_s$}\\
    \frac{2}{w_s^2} \omega & \text{$|\omega|\le w_s$}\\
    \frac{2}{w_s^2} (\omega - \sqrt{\omega^2-w_s^2}) &\text{$\omega > w_s$}\\
  \end{cases}
\end{align}

Finally, the binding energy of the impurity is defined as
\begin{align}
  E = \int_{-\infty}^{\omega=0}D_a(\omega)d\omega - \epsilon_a\:,
\end{align}
and it can be calculated in straight forward manner, in terms of the semi-circular DOS.

\section{DFT binding energies for 4$d$ metals}\label{DFTbindenergy}

\begin{figure}[h] %[!htb] %fig-2
\begin{center}
\includegraphics[width=0.45\textwidth]{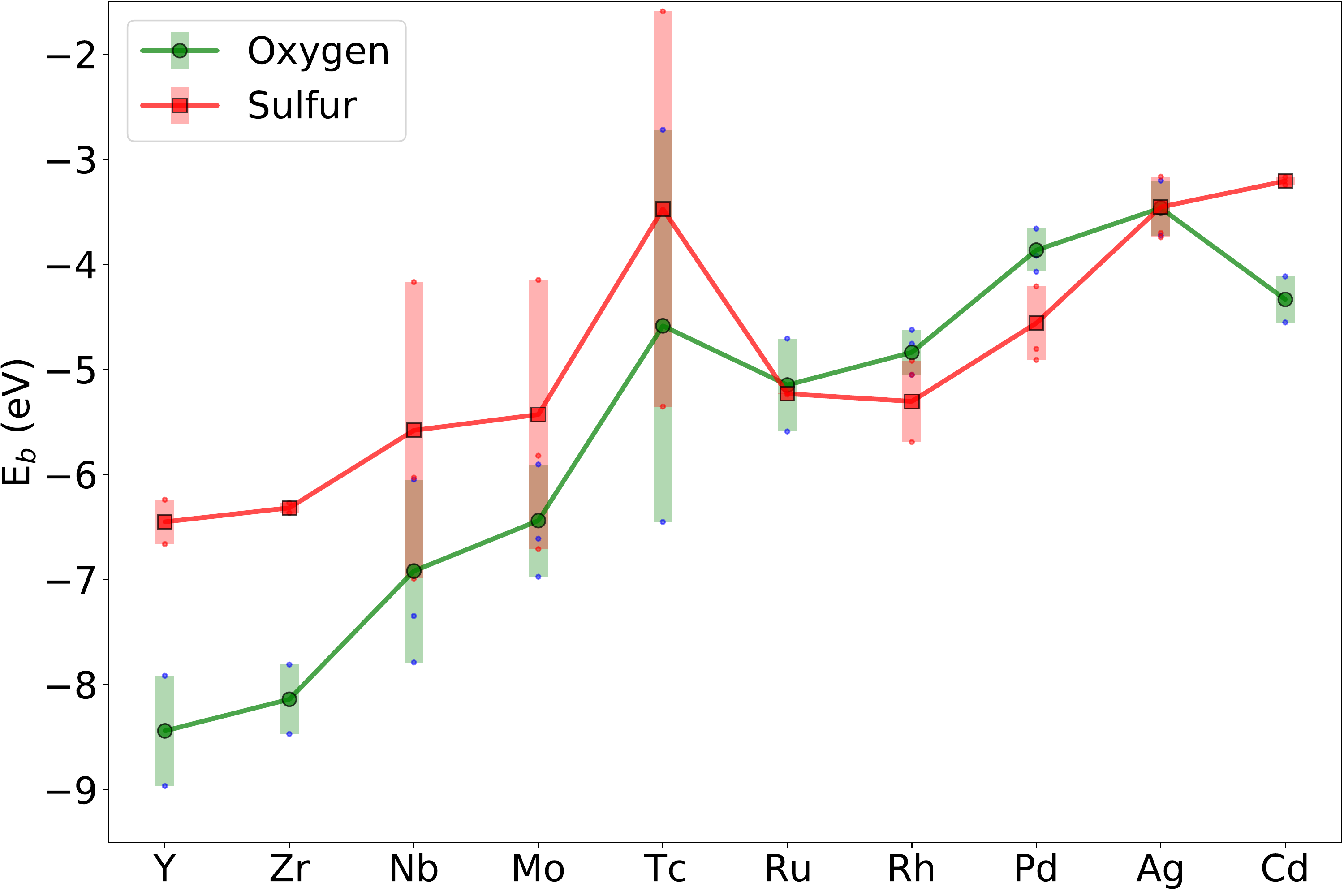}
\end{center}
\caption{(Color on line) Calculated binding energy for O and S over the 4$d$ transition-metal series. The dots represent the 
average binding energy and the bar the range of values found.}
\label{fig:fig10}
\end{figure}

In Fig.~\ref{fig:fig10} we re-plot the distribution of DFT binding energy across the 4$d$ transition-metal space as a 
function of the atomic number. In particular, for each element of the 4$d$ transition-metal series, the figure reports 
the average binding energy and its variance, when these are taken over the different surface orientations of both 
the actual structures and the hypothetical {\it fcc} ones. 

\section{Binding energies for ternary compounds}
\newcommand*{\myalign}[2]{\multicolumn{1}{#1}{#2}}
\begin{table}[h]
  \caption{\label{tab:tern-os} Composition-averaged binding energy, $\tilde{E}_\mathrm{b}$, and elemental energy shift, 
  $\Delta E^n_\mathrm{el}$ ($n=A, B, C$), for all elements in the ternary compounds $A_xB_yC_z$ investigated. All 
  energies are in eV.}
  \centering
  \resizebox{0.49\textwidth}{!}{
  \begin{tabular}{c|r|r|r|r|r|r|r|r}
    \toprule
    \hline \hline
    {} & \multicolumn{4}{c|}{Oxygen} & \multicolumn{4}{c}{Sulfur} \\
    \hline
    A$_x$B$_y$C$_z$ &  \myalign{c|}{$\tilde{E}_\mathrm{b}$} & $\Delta E_\mathrm{b}^A$ & $\Delta E_\mathrm{b}^B$ & $\Delta E_\mathrm{b}^C$ & \myalign{c|}{$\tilde{E}_\mathrm{b}$} & $\Delta E_\mathrm{b}^A$ & $\Delta E_\mathrm{b}^B$ & $\Delta E_\mathrm{b}^C$ \\
    \hline
    \midrule
        Ru$_{2}$YZn$_{20}$ & -3.91 &       2.29 &      -0.45 &      -0.00 & -3.84 &       1.99 &      -1.12 &      -0.00 \\
        FeMo$_{4}$Zr$_{9}$ & -6.90 &       1.60 &       0.72 &      -0.21 & -6.05 &       1.42 &       0.29 &      -0.94 \\
                    PtScZn & -4.92 &       0.66 &       0.21 &       0.00 & -4.28 &       0.60 &       0.23 &       0.00 \\
        Mo$_{4}$NiZr$_{9}$ & -6.94 &       0.66 &       0.06 &      -0.28 & -6.06 &       0.25 &       0.05 &      -0.97 \\
        CoMo$_{4}$Zr$_{9}$ & -6.63 &       0.35 &       0.77 &       0.08 & -5.90 &       0.33 &       0.33 &      -0.80 \\
        CoHf$_{9}$Re$_{4}$ & -7.20 &       0.35 &      -0.89 &       0.20 & -6.15 &       0.33 &      -1.33 &      -0.01 \\
        CoHf$_{9}$Mo$_{4}$ & -7.31 &       0.35 &      -0.97 &       0.01 & -6.17 &       0.33 &      -1.34 &      -0.10 \\
         CoHf$_{9}$W$_{4}$ & -7.40 &       0.35 &      -1.04 &      -0.14 & -6.15 &       0.33 &      -1.30 &      -0.49 \\
         Fe$_{4}$Mn$_{8}$Y & -5.96 &       0.24 &       0.72 &       1.40 & -5.56 &       0.23 &       0.62 &       0.83 \\
                     PdYZn & -5.10 &       0.11 &       0.26 &       0.00 & -4.41 &       0.10 &      -0.13 &       0.00 \\
                    PdScZn & -4.97 &       0.11 &       0.06 &       0.00 & -4.32 &       0.10 &       0.11 &       0.00 \\
                     PdYZn & -5.25 &       0.11 &      -0.22 &       0.00 & -4.41 &       0.10 &      -0.13 &       0.00 \\
               MnNi$_{4}$Y & -4.69 &       0.06 &       0.06 &       1.43 & -4.32 &       0.06 &       0.05 &       0.83 \\
 Ni$_{9}$Pt$_{4}$Ti$_{11}$ & -5.46 &       0.06 &       0.66 &      -0.92 & -4.72 &       0.05 &       0.60 &      -0.72 \\
              Ni$_{4}$ZnZr & -4.19 &       0.06 &       0.00 &       0.61 & -3.93 &       0.05 &       0.00 &       0.37 \\
        Ni$_{9}$Ti$_{2}$Zr & -4.60 &       0.06 &      -1.01 &       0.54 & -4.20 &       0.05 &      -0.81 &       0.27 \\
               CdRuY$_{4}$ & -7.19 &       0.00 &       2.29 &      -0.72 & -5.91 &       0.00 &       1.99 &      -1.28 \\
              AuLuNi$_{4}$ & -4.16 &       0.00 &       1.07 &       0.06 & -3.87 &       0.00 &       0.56 &       0.05 \\
              CdLuNi$_{4}$ & -4.18 &       0.00 &       0.99 &       0.06 & -3.88 &       0.00 &       0.51 &       0.05 \\
              CdPt$_{2}$Zn & -3.70 &       0.00 &       0.66 &       0.00 & -3.70 &       0.00 &       0.60 &       0.00 \\
               AuNi$_{4}$Y & -4.23 &       0.00 &       0.06 &       1.27 & -3.91 &       0.00 &       0.05 &       0.77 \\
               CdNi$_{4}$Y & -4.25 &       0.00 &       0.06 &       1.13 & -3.92 &       0.00 &       0.05 &       0.67 \\
              AuNi$_{4}$Sc & -4.26 &       0.00 &       0.06 &       0.50 & -3.95 &       0.00 &       0.05 &       0.46 \\
              CdNi$_{4}$Sc & -4.33 &       0.00 &       0.06 &       0.06 & -4.01 &       0.00 &       0.05 &       0.11 \\
              Cu$_{2}$NiZn & -3.70 &       0.00 &       0.06 &       0.00 & -3.70 &       0.00 &       0.05 &       0.00 \\
              CuNi$_{2}$Ti & -4.63 &       0.00 &       0.06 &      -0.83 & -4.22 &       0.00 &       0.05 &      -0.63 \\
              Cu$_{2}$ZnZr & -4.46 &       0.00 &       0.00 &       0.55 & -4.06 &       0.00 &       0.00 &       0.32 \\
              CdCu$_{2}$Zr & -4.47 &       0.00 &       0.00 &       0.49 & -4.06 &       0.00 &       0.00 &       0.29 \\
              CdHg$_{2}$Ti & -4.39 &       0.00 &       0.00 &       0.13 & -4.07 &       0.00 &       0.00 &      -0.01 \\
              AgAuCd$_{2}$ & -3.70 &       0.00 &       0.00 &       0.00 & -3.70 &       0.00 &       0.00 &       0.00 \\
              AgAuZn$_{2}$ & -3.70 &       0.00 &       0.00 &       0.00 & -3.70 &       0.00 &       0.00 &       0.00 \\
              AuCuZn$_{2}$ & -3.70 &       0.00 &       0.00 &       0.00 & -3.70 &       0.00 &       0.00 &       0.00 \\
              AgHg$_{2}$Ti & -4.43 &       0.00 &       0.00 &      -0.03 & -4.13 &       0.00 &       0.00 &      -0.26 \\
 Cu$_{7}$Zn$_{16}$Zr$_{6}$ & -4.60 &       0.00 &       0.00 &      -0.77 & -4.27 &       0.00 &       0.00 &      -1.02 \\
              CuHg$_{2}$Ti & -4.84 &       0.00 &       0.00 &      -1.67 & -4.47 &       0.00 &       0.00 &      -1.61 \\
              CuHg$_{2}$Zr & -5.15 &       0.00 &       0.00 &      -2.22 & -4.59 &       0.00 &       0.00 &      -1.80 \\
              CuFePt$_{2}$ & -4.37 &       0.00 &      -0.38 &       0.66 & -4.31 &       0.00 &      -0.41 &       0.60 \\
              CuHfHg$_{2}$ & -4.70 &       0.00 &      -0.78 &       0.00 & -4.24 &       0.00 &      -0.74 &       0.00 \\
 Cu$_{7}$Hf$_{6}$Zn$_{16}$ & -4.62 &       0.00 &      -1.25 &       0.00 & -4.24 &       0.00 &      -1.21 &       0.00 \\
        FeHf$_{9}$Re$_{4}$ & -7.42 &      -0.07 &      -0.94 &       0.15 & -6.30 &       0.02 &      -1.33 &      -0.03 \\
        FeHf$_{9}$Mo$_{4}$ & -7.40 &      -0.07 &      -0.89 &       0.03 & -6.31 &       0.02 &      -1.35 &      -0.10 \\
         FeHf$_{9}$W$_{4}$ & -7.47 &      -0.11 &      -0.93 &      -0.11 & -6.31 &      -0.01 &      -1.31 &      -0.50 \\
         RuV$_{3}$Zr$_{2}$ & -6.67 &      -0.24 &       0.02 &      -0.00 & -5.47 &      -0.16 &       0.03 &       0.03 \\
              FeNiPt$_{2}$ & -4.55 &      -0.34 &      -0.74 &       0.66 & -4.49 &      -0.38 &      -0.69 &       0.60 \\
         Cr$_{4}$Fe$_{8}$Y & -6.00 &      -0.37 &       0.31 &       1.11 & -5.55 &      -0.59 &       0.30 &       0.74 \\
              CoCrPt$_{2}$ & -4.65 &      -0.76 &      -0.38 &       0.66 & -4.45 &      -0.65 &      -0.38 &       0.60 \\
        Hf$_{9}$NiRe$_{4}$ & -7.20 &      -0.88 &       0.06 &       0.16 & -6.15 &      -1.31 &       0.05 &      -0.03 \\
  Co$_{12}$Mn$_{5}$Y$_{2}$ & -5.74 &      -0.97 &       0.20 &       0.04 & -5.27 &      -0.86 &       0.26 &       0.04 \\
        Hf$_{9}$Mo$_{4}$Ni & -7.34 &      -1.02 &       0.02 &       0.06 & -6.17 &      -1.36 &      -0.10 &       0.05 \\
    Hf$_{9}$NiW$_{4}$ & -7.45 &      -1.09 &       0.06 &      -0.13 & -6.16 &      -1.30 &       0.05 &      -0.49 \\
    \bottomrule
    \hline\hline
  \end{tabular}
  }
  \end{table}

\section{Tables with thermodynamical parameters for oxides and sulphides}
\label{materials_tables}

%\begin{table}
%\caption{\label{tab:4d-2} $EN$ is the Pauling electronegativity, }
%\begin{tabular}{lcccccccc} \hline\hline
%El. &  $d_\mathrm{X-O}^\mathrm{LDA}$  & $n_\mathrm{X-O}^\mathrm{LDA}$ & $d_\mathrm{X-O}^\mathrm{RPBE}$  & $n^\mathrm{RPBE}_\mathrm{X-O}$ &
%$d_\mathrm{X-S}^\mathrm{LDA}$  & $n_\mathrm{X-S}^\mathrm{LDA}$ & $d_\mathrm{X-S}^\mathrm{RPBE}$  & $n^\mathrm{RPBE}_\mathrm{X-S}$ \\
%\hline
%Y   & 2.13 & 3 & 2.18 & 3 & 2.62 & 3 & 2.69 & 3  \\
%Zr  & 2.05 & 3 & 2.09 & 3 & 2.52 & 3 & 2.58 & 3  \\
%Nb & 2.18 & 1 & 2.26 & 1 & 2.55 & 5 & 2.63 & 4  \\
%Mo & 2.08 & 1 & 2.16 & 1 & 2.48 & 5 & 2.52 & 4  \\
%Tc  & 2.01 & 3 & 2.05 & 3 & 2.35 & 3 & 2.40 & 3  \\
%Ru & 2.02 & 3 & 2.06 & 3 & 2.33 & 3 & 2.37 & 3  \\
%Rh & 2.11 & 4 & 2.15 & 4 & 2.32 & 4 & 2.36 & 4  \\
%Pd & 2.12 & 4 & 2.16 & 4 & 2.29 & 4 & 2.34 & 4  \\
%Ag & 2.21 & 4 & 2.27 & 4 & 2.48 & 4 & 2.56 & 4  \\
%Cd & 2.07 & 3 & 2.13 & 3 & 2.48 & 3 & 2.50 & 3  \\ \hline\hline
%\end{tabular}
%\end{table}

\begin{table*}
\caption{\label{tab:3dO} Summary table of structural information and enthalpy of formation, $\Delta H_f$, for 3$d$ 
transition-metal oxides. Note that there are compounds for which the information is incomplete. Data from this table
has been included in Fig.~\ref{fig:1}.}
\begin{tabular}{p{1.3cm}p{1.9cm}p{1.8cm}p{3.4cm}p{2.9cm}p{2.4cm}p{1.0cm}} \hline\hline
$Z$ & Compound & SG & Lattice Constants (\AA) & $\Delta H_f$ (kcal mol$^{-1}$) & $\Delta H_f$/atom (eV) & Ref. \\
\hline\hline
21 Sc			&  Sc$_2$O$_3$ &  $Ia\bar{3}$     & 9.79, 9.79, 9.79          & 456 & 3.96 & [\onlinecite{Sc2O3stru},\onlinecite{Curta}] \\ \hline
22 Ti 			  &     Ti$_6$O     &  $P\bar{3}1c$  &  5.13, 5.13, 9.48      &  &  & [\onlinecite{Ti3Ostru}] \\
			          &     Ti$_3$O     & $P\bar{3}1c$   &  5.15, 5.15, 9.56      &  &  & [\onlinecite{Ti3Ostru}] \\
			          &     Ti$_2$O     & $P\bar{3}m1$   &  2.9194, 2.9194, 4.713      &  &  & [\onlinecite{Novoselova2004}] \\
			          &     Ti$_3$O$_2$ & $P6/mmm$   &        &  &  & [\onlinecite{CRC}] \\
			          &     TiO     &  	$P\bar{6}m2$  & 3.031, 3.031, 3.2377       & 129.5 & 2.81 & [\onlinecite{TiOstru},\onlinecite{Curta}] \\
			          &      Ti$_2$O$_3$    &  $R\bar{3}c$  &  5.126, 5.126, 13.878      & 363.29 & 3.15 & [\onlinecite{Ti2O3stru},\onlinecite{Curta}] \\
			          &      Ti$_3$O$_5$    & $C2/m$   &   9.752, 3.802, 9.442     & 587.72 & 3.19 & [\onlinecite{Ti3O5stru},\onlinecite{Curta}] \\
			          &      TiO$_2$    & $P4_2/mnm$   &  4.6257, 4.6257, 2.9806    & 225.6 & 3.26 & [\onlinecite{TiO2Rstru},\onlinecite{Curta}] \\
			          &      TiO$_2$    &  $I4_1/amd$  &   3.771, 3.771, 9.43     & 224.9 & 3.25 &  [\onlinecite{TiO2Astru},\onlinecite{Curta}] \\  \hline
23 V				  &      VO    &  	$Fm\bar{3}m$  &    4.0678, 4.0678, 4.0678    & 170.60 & 3.7 & [\onlinecite{VOstru},\onlinecite{Curta}] \\
			          &      V$_5$O$_9$ & $P\bar{1}$   & 7.005, 8.3629, 10.9833          &  &  & [\onlinecite{V5O9}]  \\
			          &      V$_4$O$_7$ & $P\bar{1}$   & 5.504, 7.007, 10.243          &  &  & [\onlinecite{V4O7Horiuchi}]  \\
			          &      V$_3$O$_5$    &  $Cc$  &    9.98, 5.03, 9.84    &  &  & [\onlinecite{V3O5}] \\
			          &      V$_2$O$_3$    &  $R\bar{3}c$  &    4.9776, 4.9776, 13.9647    & 291.3 & 2.53 & [\onlinecite{V2O3stru},\onlinecite{Curta}] \\
			          &      VO$_2$    & $P4_2/mnm$   &  4.53, 4.53, 2.869      & 170.6 & 2.47 & [\onlinecite{VO2stru},\onlinecite{Curta}] \\
			          &      V$_6$O$_{13}$    & $C2/m$   &  11.922, 3.68, 10.138      &  &  & [\onlinecite{V6O13}] \\
			          &      V$_2$O$_{5}$    & $Pmn2_1$   &  11.48, 4.36, 3.555      & 370.46 & 2.3 & 		[\onlinecite{V2O5stru},\onlinecite{Curta}] \\ \hline
24 Cr		       &    Cr$_2$O$_3$      &  $R\bar{3}c$   &   4.9607, 4.9607, 13.599     & 271.2 & 2.35 & [\onlinecite{Cr2O3stru},\onlinecite{Curta}] \\
			       &    Cr$_3$O$_4$      &     &        & 365.92 & 2.27 & [\onlinecite{CRC}] \\
			       &    CrO$_2$      &  $P4_2/mnm$   &   4.421, 4.421, 2.917     & 142.9 & 2.07 &  [\onlinecite{CrO2stru},\onlinecite{CRC}] \\
			       &    CrO$_3$      &  $Ama2$   &   5.743, 8.557, 4.789     & 140.3 & 1.52 & [\onlinecite{CrO3stru},\onlinecite{Curta}] \\ \hline
25 Mn		  &    MnO      &  $Fm\bar{3}m$   &    4.444, 4.444, 4.444    & 91.99 & 1.99 & [\onlinecite{MnOstru},\onlinecite{Curta}] \\
			  &    Mn$_3$O$_4$      & $I4_1/amd$   &   5.76, 5.76, 9.46     & 331.6 & 2.05 & [\onlinecite{Mn3O4stru},\onlinecite{Curta}]  \\
			  &    Mn$_2$O$_3$      &  $Ia\bar{3}$   & 9.42, 9.42, 9.42       & 229.00  &1.99  & [\onlinecite{Mn2O3stru},\onlinecite{Curta}] \\
			  &    MnO$_2$      & $P4_2/mnm$    & 4.4, 4.4, 2.87       & 125.5  &1.81  & [\onlinecite{MnO2stru},\onlinecite{NewBook}] \\ \hline
26 Fe 		       &   FeO       &  $Fm\bar{3}m$   &    4.303, 4.303, 4.303    & 196.8 & 1.71 & [\onlinecite{FeOstru},\onlinecite{Curta}] \\
			       &   Fe$_2$O$_3$       &   $R\bar{3}c$  &  5.43, 5.43, 5.43      & 196.8 & 1.71 & [\onlinecite{Fe2O3stru},\onlinecite{Curta}] \\
			       &   Fe$_3$O$_4$       &   $Fd\bar{3}m$  &   8.3965, 8.3965, 8.3965     & 265.01 & 1.64 & [\onlinecite{Fe3O4stru},\onlinecite{Curta}] \\ \hline
27 Co		  &    CoO      &  $Fm\bar{3}m$   &    4.258, 4.258, 4.258    & 56.81 & 1.23 & [\onlinecite{CoOstru},\onlinecite{Curta}] \\
			  &    Co$_3$O$_4$      &  $Fd\bar{3}m$   &  8.0821, 8.0821, 8.0821      & 217.5 & 1.35 & [\onlinecite{Co3O4stru},\onlinecite{Curta}] \\ \hline
28 Ni				  &   NiO       &  $Fm\bar{3}m$   &   4.1684, 4.1684, 4.1684     & 57.29 & 1.24 & [\onlinecite{NiOstru},\onlinecite{Curta}] \\ \hline
29 Cu    			  &    Cu$_2$O      & $Pn\bar{3}m$    &  4.252, 4.252, 4.252      & 41.39 & 0.6 & [\onlinecite{Cu2Ostru},\onlinecite{Curta}]  \\
	    			  &    CuO      &  $P2_1/c$   &  4.683, 3.4203, 5.1245      & 38.69 & 0.84 & [\onlinecite{CuOstru},\onlinecite{Curta}] \\ \hline
30 Zn		  &    ZnO      &  $P6_3mc$  &  3.249, 3.249, 5.207      & 83.77 & 1.82 & [\onlinecite{ZnOstru},\onlinecite{Curta}]   \\ \hline\hline
\end{tabular}
\end{table*}

\begin{table*}
\caption{\label{tab:3dS} Summary table of structural information and enthalpy of formation, $\Delta H_f$, for 3$d$ 
transition-metal sulphides. Note that there are compounds for which the information is incomplete. Data from this table
has been included in Fig.~\ref{fig:1}.}
\begin{tabular}{p{1.3cm}p{1.9cm}p{1.8cm}p{3.4cm}p{2.9cm}p{2.4cm}p{1.0cm}} \hline\hline
$Z$ & Compound & SG & Lattice Constants (\AA) & $\Delta H_f$ (kcal mol$^{-1}$) & $\Delta H_f$/atom (eV) & Ref. \\
\hline\hline
21 Sc		          &  ScS &  $Fm\bar{3}m$     & 5.166, 5.166, 5.166          & 57.36 & 1.24 & [\onlinecite{ScSstru},\onlinecite{ScSthe}] \\ \hline
22 Ti 			  &     Ti$_6$S     &     &        &  &  & [\onlinecite{STiSystem}] \\ %\onlinecite{XXX}
			          &    Ti$_3$S      &    &        &  &  & [\onlinecite{STiSystem}] \\
			          &    Ti$_8$S$_3$  & $C2/m$   &  32.69, 3.327, 19.36  &         &  & [\onlinecite{Ti8S3stru}] \\
			          &    Ti$_2$S      &    &        &  &  & [\onlinecite{STiSystem}] \\
			          &   TiS       &  	$P6_{3}/mmc$  &    3.299, 3.299, 6.38    & 64.5 & 1.4 & [\onlinecite{TiSstru},\onlinecite{Curta}] \\
			          &    Ti$_4$S$_5$  &	 $P6_{3}/mmc$    & 3.439, 3.439, 28.933   &        &  & [\onlinecite{Ti4S5stru}]  \\
			          &    Ti$_3$S$_4$      &  $P6_{3}/mmc$  &   3.43, 3.43, 11.4     &  &  & [\onlinecite{Ti3S4stru}]  \\
			          &    Ti$_2$S$_3$      & $P6_3/mc$   & 3.422, 3.422, 11.442 & 147.7 & 1.28 & [\onlinecite{Ti2S3stru},\onlinecite{Ti2S3the}]  \\
			          &     TiS$_2$     & $P\bar{3}m1$   & 3.397, 3.397, 5.691       & 97.3 & 1.41 & [\onlinecite{TiS2stru},\onlinecite{Curta}] \\
			          &      TiS$_3$    & $P2_1/m$   &   4.9476, 3.3787, 8.7479     & 100.1 & 1.09 & [\onlinecite{TiS3stru},\onlinecite{Ti2S3the}] \\  \hline
23 V			  &   V$_3$S       &  $I\bar{4}m2$   &   9.47, 9.47, 4.589     &  &  & [\onlinecite{V3Sstru}] \\
			          &      V$_5$S$_4$    & $I4/m$   & 8.988, 8.988, 3.224       &  &  & [\onlinecite{V5S4stru}] \\
			          &      VS    &  $P6_{3}/mmc$  &   3.34, 3.34, 5.785     &  &  &  [\onlinecite{VSstru}] \\
			          &      V$_7$S$_8$    & $	P3_221$   & 6.706, 6.706, 17.412       &  &  & [\onlinecite{V5S4stru}] \\
			          &      V$_3$S$_4$    & $C2/m$   &   12.599, 3.282, 5.867     &  &  & [\onlinecite{V3S4stru}] \\
			          &      V$_5$S$_8$    & $C2/m$   & 11.3, 6.6, 8.1       &  &  & [\onlinecite{V3S4stru}] \\
			          &      VS$_4$    &  $I2/a$  &  6.78, 10.42, 12.11      &  &  & [\onlinecite{VS4stru}] \\ \hline
24 Cr			  &   CrS       &  $C2/c$   &    3.826, 5.913, 6.089    & 37.19 & 0.81 & [\onlinecite{CrSstru},\onlinecite{Curta}] \\
			  &   Cr$_2$S$_3$       &  $R\bar{3}$   &    5.937, 5.937, 16.698    &  &  & [\onlinecite{CrSstru},\onlinecite{Curta}] \\ \hline
25 Mn			  &    MnS      & $Fm\bar{3}m$    &   5.24, 5.24, 5.24     & 51.19 & 1.11 & [\onlinecite{MnSstru},\onlinecite{Curta}]  \\
			  &    MnS$_2$      &  $Pa\bar{3}$   &   6.083, 6.083, 6.083     & 49.50 & 0.72 &[\onlinecite{MnS2stru},\onlinecite{Curta}] \\ \hline
26 Fe			  &   FeS       & $P6_3/mmc$    &   3.445, 3.445, 5.763     & 24.0 & 0.52 & [\onlinecite{FeSstru},\onlinecite{Curta}] \\
			       &   Fe$_3$S$_4$       &   $Fd\bar{3}m$  &    9.876, 9.876, 9.876    &  &  & [\onlinecite{Fe3S4stru},\onlinecite{Curta}] \\
			       &   FeS$_2$       & $Pa\bar{3}$    &  5.4179, 5.4179, 5.4179      & 40.99 & 0.59 &  [\onlinecite{FeS2stru},\onlinecite{Curta}] \\ \hline
27 Co			  &   Co$_9$S$_8$       &  $Fm\bar{3}m$   &   9.927, 9.927, 9.927     & 22.61 & 0.49 & [\onlinecite{Co9S8stru},\onlinecite{Curta}] \\
			  &   Co$_3$S$_4$       &  $Fd\bar{3}m$   &  9.401, 9.401, 9.401      & 85.8 & 0.53 & [\onlinecite{Co3S4stru},\onlinecite{Curta}]  \\
			  &    CoS$_2$       &   $Pa\bar{3}$  &   5.5385, 5.5385, 5.5385     & 36.59 & 0.53 & [\onlinecite{CoS2stru},\onlinecite{Curta}] \\ \hline
28 Ni			  &   Ni$_3$S$_2$       &  $R32$   &   4.049, 4.049, 4.049     & 51.70 & 0.45 & [\onlinecite{Ni3S2stru},\onlinecite{Curta}] \\
			  &   NiS       &   $P6_3mc$  &  3.4456, 3.4456, 5.405     & 23.4 & 0.51 & [\onlinecite{NiSstru},\onlinecite{Curta}] \\
			  &   Ni$_3$S$_4$       &  $Fd\bar{3}m$   &  9.65, 9.65, 9.65      & 71.99 &0.45  & [\onlinecite{Wyckoff2},\onlinecite{Curta}] \\
			  &   NiS$_2$       &  $Pa\bar{3}$   &   5.6873, 5.6873, 5.6873     & 29.85 & 0.43 & [\onlinecite{NiS2stru},\onlinecite{NiBook}] \\ \hline
29 Cu			  &    Cu$_2$S      &  $Pa\bar{3}$   & 5.7891, 5.7891, 5.7891        & 19.0 & 0.27 & [\onlinecite{Cu2Sstru},\onlinecite{Curta}]  \\
			      &    CuS      &  $P6_3/mmc$   &    3.7938, 3.7938, 16.341    & 12.5 & 0.27 & [\onlinecite{CuSstru},\onlinecite{Curta}] \\ \hline
30 Zn			  &   ZnS       & 	$P6_3mc$    &   3.8227, 3.8227, 6.2607     & 49.0 & 1.06 & [\onlinecite{ZnSstru},\onlinecite{Curta}] \\ \hline\hline
\end{tabular}
\end{table*}

\begin{table*}
\caption{\label{tab:4dO} Summary table of structural information and enthalpy of formation, $\Delta H_f$, for 4$d$ 
transition-metal oxides. Note that there are compounds for which the information is incomplete. Data from this table
has been included in Fig.~\ref{fig:1}.}
\begin{tabular}{p{1.3cm}p{1.6cm}p{3.0cm}p{3.4cm}p{2.9cm}p{2.0cm}p{1.0cm}} \hline\hline
$Z$ & Compound & SG & Lattice Constants (\AA) & $\Delta H_f$ (kcal mol$^{-1}$) & $\Delta H_f$ (eV) & Ref. \\
\hline\hline
39 Y			          &   Y$_2$O$_3$       & $Ia\bar{3}$   & 10.596, 10.596, 10.596       & 455.37 & 3.95 & [\onlinecite{Y2O3stru},\onlinecite{Curta}] \\ \hline
40 Zr			          &    ZrO$_2$      & $P2_1/c$   & 5.1462, 5.2082, 5.3155       & 262.9 & 3.8 & [\onlinecite{ZrO2Mstru},\onlinecite{Curta}] \\
			          &    ZrO$_2$      & $P4_2/nmc$  &   3.5781, 3.5781, 5.1623     & 262.9 & 3.8 & [\onlinecite{ZrO2Tstru},\onlinecite{Curta}] \\
			          &    ZrO$_2$      &  $Fm\bar{3}m$  & 5.1291, 5.1291, 5.1291       & 262.9 & 3.8 & [\onlinecite{ZrO2Cstru},\onlinecite{Curta}] \\ \hline
41 Nb		         & NbO         & $Pm\bar{3}m$   & 4.2, 4.2, 4.2       & 100.31 & 2.17 & [\onlinecite{NbOstru},\onlinecite{Curta}] \\
			         & NbO$_2$         &  $I4_1/a$ &  13.66, 13.66, 5.964      & 190.30 & 2.75 & [\onlinecite{NbO2stru},\onlinecite{CRC}] \\
			         & Nb$_2$O$_5$         & $I4/mmm$ &  20.44, 20.44, 3.832      & 454.00 & 2.81 & [\onlinecite{Nb2O5stru},\onlinecite{Curta}] \\ \hline
42 Mo		          &  MoO$_2$        & $P2_1/c$   & 5.584, 4.842, 5.608     & 140.51 & 2.03 & [\onlinecite{Wyckoff},\onlinecite{Curta}] \\
			          &  MoO$_3$        &   $Pnma$ & 13.825, 3.694, 3.954       & 178.11 & 1.93 & [\onlinecite{MoO3stru},\onlinecite{Curta}] \\ \hline
43 Tc		          &    TcO$_2$      &  $P1_2/c$   &   5.6891, 4.7546, 5.5195     & 109.42 & 1.58 & [\onlinecite{TcO2stru},\onlinecite{TcS2thermo}] \\
			          &    Tc$_2$O$_7$      & $Pbca$    &   13.756, 7.439, 5.617     & 269.24 & 1.30 & [\onlinecite{Tc2O7stru},\onlinecite{TcS2thermo}] \\ \hline
44 Ru		          &    RuO$_2$      &  $P4_2/mnm$  & 4.4968, 4.4968, 3.1049       & 72.9 & 1.05 & [\onlinecite{RuO2stru},\onlinecite{Curta}] \\ \hline
45 Rh		          &    Rh$_2$O$_3$      & $Pbca$   & 5.1477, 5.4425, 14.6977  & 84.99 & 0.74 & [\onlinecite{Rh2O3stru},\onlinecite{Curta}] \\ \hline
46 Pd		  &      PdO    & $P4_2/mmc$ &  3.03, 3.03, 5.33      & 27.61 & 0.60 & [\onlinecite{PdOstru},\onlinecite{Curta}] \\ \hline
47 Ag		          &    Ag$_2$O      & 	$Pn\bar{3}m$    & 4.7306, 4.7306, 4.7306  & 7.43 & 0.11 & [\onlinecite{Ag2Ostru},\onlinecite{Curta}]  \\ \hline
48 Cd		  &    CdO      & $Fm\bar{3}m$ & 4.699, 4.699, 4.699  & 61.76 & 1.34 & [\onlinecite{CdOstru},\onlinecite{Curta}]  \\ \hline\hline
\end{tabular}
\end{table*}

\begin{table*}
\caption{\label{tab:4dS} Summary table of structural information and enthalpy of formation, $\Delta H_f$, for 4$d$ 
transition-metal sulphides. Note that there are compounds for which the information is incomplete. Data from this table
has been included in Fig.~\ref{fig:1}.}
\begin{tabular}{p{1.3cm}p{1.6cm}p{3.0cm}p{3.4cm}p{2.9cm}p{2.0cm}p{1.0cm}} \hline\hline
$Z$ & Compound & SG & Lattice Constants (\AA) & $\Delta H_f$ (kcal mol$^{-1}$) & $\Delta H_f$ (eV) & Ref. \\
\hline\hline
39 Y   			  &   Y$_2$S$_3$       & $P2_1/m$    &   17.5234, 4.0107, 10.1736     &  &  & [\onlinecite{Y2S3stru}] \\ \hline
40 Zr 			  &   Zr$_3$S$_2$       & $P\bar{6}m2$    &  3.429, 3.429, 3.428      & 88.34 & 0.77 & [\onlinecite{ZrBook}] \\
	 			  &   ZrS       &  $Fm\bar{3}m$   &    5.25, 5.25, 5.25    &  &  & [\onlinecite{Wyckoff},\onlinecite{Curta}] \\
	 			  &   ZrS$_2$       &  $P\bar{3}m1$   &   3.63, 3.63, 5.85     & 138 & 1.99 & [\onlinecite{Wyckoff2},\onlinecite{Curta}] \\
	 			  &   ZrS$_3$       & $P2_1/m$    & 5.1243, 3.6244, 8.980       & 148.11 & 1.61 & [\onlinecite{ZrS3stru},\onlinecite{ZrBook}] \\ \hline
41 Nb			    &   NbS       &   $P6_3/mmc$  &  3.32, 3.32, 6.46      &  &  & [\onlinecite{NbSstru}] \\
				    &   NbS$_2$       & $P\bar{6}2c$    &  3.35, 3.35, 17.94      &  &  & [\onlinecite{NbS2Astru}] \\
				    &   NbS$_2$       & $P\bar{3}m1$    &  3.42, 3.42, 5.938      &  &  & [\onlinecite{NbS2Bstru}] \\ \hline
42 Mo			  &    Mo$_2$S$_3$      & $P2_1/m$   & 6.092, 3.208, 8.6335       & 97.20 & 0.84 & [\onlinecite{Mo2S3stru},\onlinecite{Curta}] \\
			          &     MoS$_2$     & $P6_3/mmc$ &  3.169, 3.169, 12.324      & 65.89 & 0.95 & [\onlinecite{MoS2stru},\onlinecite{Curta}] \\ \hline
43 Tc			  &   TcS$_2$       &  P1   &   6.456, 6,357, 6.659     & 53.49 & 0.77 & [\onlinecite{TcS2stru},\onlinecite{TcS2thermo}] \\
			          &   Tc$_2$S$_7$       & unknown    &        & 147 & 0.71 & [\onlinecite{TcS2thermo}] \\ \hline
44 Ru			  &    RuS$_2$      &  $Pa\bar{3}$   & 5.6106, 5.6106, 5.6106       & 49.21 & 0.71 & [\onlinecite{RuS2stru},\onlinecite{Curta}] \\ \hline
45 Rh			  &   Rh$_3$S$_4$       & $C2/m$    &  10.4616, 10.7527, 6.2648  & 84.54 & 0.52 & [\onlinecite{Rh3S4stru},\onlinecite{Curta}] \\
			          &    Rh$_2$S$_3$      & $Pbcn$ &  8.462, 5.985, 6.138      & 62.81 & 0.54 & [\onlinecite{Rh2S3stru},\onlinecite{Curta}] \\
		              &   RhS$_2$       & $Pa\bar{3}$ &  5.57, 5.57, 5.57      &  &  & [\onlinecite{RhS2stru},\onlinecite{Curta}] \\ \hline
46 Pd		  &     Pd$_4$S     & $P\bar{4}2_1c$    &   5.1147, 5.1147, 5.5903  & 16.5 & 0.14 & [\onlinecite{Pd4Sstru},\onlinecite{Curta}] \\
			  &      PdS    & $P4_2/m$ & 6.429, 6.429, 6.611       & 16.90 & 0.37 & [\onlinecite{PdSstru},\onlinecite{Curta}] \\
			  &      PdS$_2$    & $Pbca$   & 5.46, 5.541, 7.531       & 18.69 & 0.27 & [\onlinecite{PdS2stru},\onlinecite{Curta}] \\ \hline
47 Ag			  &  Ag$_2$S        &  $P2_1/m$   & 4.229, 6.931, 7.862 & 7.60 & 0.11 & [\onlinecite{Ag2Sstru},\onlinecite{Curta}]  \\ \hline
48 Cd			  &   CdS       & $P6_3mc$ &  4.137, 4.137, 6.7144      & 35.70 & 0.77 & [\onlinecite{ZnOstru},\onlinecite{Curta}]  \\ \hline
\end{tabular}
\end{table*}

\begin{table*}
\caption{\label{tab:5dO} Summary table of structural information and enthalpy of formation, $\Delta H_f$, for 5$d$ 
transition-metal oxides. Note that there are compounds for which the information is incomplete. Data from this table
has been included in Fig.~\ref{fig:1}.}
\begin{tabular}{p{1.3cm}p{1.6cm}p{3.0cm}p{3.4cm}p{2.9cm}p{2.0cm}p{1.0cm}} \hline\hline
$Z$ & Compound & SG & Lattice Constants (\AA) & $\Delta H_f$ (kcal mol$^{-1}$) & $\Delta H_f$ (eV) & Ref. \\
\hline\hline
57 La		          &  La$_2$O$_3$        & $P6_3/mmc$   &  4.057, 4.057, 6.43      & 429 & 3.72 & [\onlinecite{La2O3stru},\onlinecite{Curta}] \\ \hline
72 Hf			  &   HfO$_2$       & $P2_1/c$    &  5.1156, 5.1722, 5.2948      & 267.09 & 3.86 & [\onlinecite{HfO2stru},\onlinecite{Curta}] \\ \hline
73 Ta		          &  Ta$_2$O$_5$        & $Pccm$   & 6.217, 3.677, 7.794       & 489.01 & 3.03 & [\onlinecite{Ta2O5stru},\onlinecite{Curta}] \\ \hline
74 W			          &      WO$_2$    & $P4_2/mnm$   & 4.86, 4.86, 2.77       & 140.89 & 2.04 & [\onlinecite{Wyckoff2},\onlinecite{Curta}] \\
			          &      W$_2$O$_5$    &    &        & 311.20 & 1.93 & [\onlinecite{W2O5stru},\onlinecite{NewBook}] \\
			          &      WO$_3$    &  $Pnma$  &   7.57, 7.341, 7.754     & 201.41 & 2.18 & [\onlinecite{WO3stru},\onlinecite{Curta}] \\ \hline
75 Re		          &     ReO$_2$     &  $Pbcn$   & 4.8094, 5.6433, 4.6007  & 103.39 & 1.49 & [\onlinecite{Wyckoff2},\onlinecite{Curta}] \\
			          &     ReO$_3$     &  $Pm\bar{3}m$  &   3.734, 3.734, 3.734     & 146.01 & 1.58 & [\onlinecite{ReO3stru},\onlinecite{Curta}] \\
			          &     Re$_2$O$_7$ & $P2_12_12_1$ & 12.508, 15.196, 5.448     & 298.40 & 1.44 & [\onlinecite{Re2O7stru},\onlinecite{Curta}] \\ \hline
76 Os		          &   OsO$_2$       & $P4_2/mnm$    &  4.519, 4.519, 3.196      & 70.41 & 1.02 & [\onlinecite{OsO2stru},\onlinecite{Curta}] \\
			          &   OsO$_4$       &  $C2$   &  8.66, 4.52, 4.75      & 94.10 & 0.81 & [\onlinecite{OsO4stru},\onlinecite{Curta}] \\ \hline
77 Ir			  &  IrO$_2$        &   $P4_2/mnm$  &  4.5051, 4.5051, 3.1586      & 59.61 & 1.29 & [\onlinecite{IrO2stru},\onlinecite{Curta}] \\ \hline
78 Pt			  &    PtO      &  $Fm\bar{3}m$   &  5.15, 5.15, 5.15      & 17 & 0.37 & [\onlinecite{PtOstru},\onlinecite{PtOenta}] \\
			  &    Pt$_3$O$_4$      &  $Im\bar{3}$   &  6.238, 6.238, 6.238     & 64.05 & 0.40 & [\onlinecite{Pt3O4stru},\onlinecite{Pt3O4enta}] \\
		  &    PtO$_2$      &  $Pnnm$   &  4.486, 4.537, 3.138      & 19.1 & 0.28 & [\onlinecite{PtO2stru},\onlinecite{PtOenta}] \\ \hline
79 Au			  &  Au$_2$O$_3$        &  $Fdd 2$   & 12.827, 10.52, 3.838       & 0.81 & 0.017 & [\onlinecite{Au2O3stru},\onlinecite{Curta}] \\ \hline
80 Hg		  &     Hg$_2$O     &   &        & 21.50 & 0.31 & [\onlinecite{NewBook}] \\
			  &     HgO     & $Pnma$   &   6.6129, 5.5208, 3.5219     & 21.70 & 0.47 & [\onlinecite{HgOstru},\onlinecite{Curta}] \\ \hline\hline
\end{tabular}
\end{table*}

\begin{table*}
\caption{\label{tab:5dS} Summary table of structural information and enthalpy of formation, $\Delta H_f$, for 5$d$ 
transition-metal sulphides. Note that there are compounds for which the information is incomplete. Data from this table
has been included in Fig.~\ref{fig:1}.}
\begin{tabular}{p{1.3cm}p{1.6cm}p{3.0cm}p{3.4cm}p{2.9cm}p{2.0cm}p{1.0cm}} \hline\hline
$Z$ & Compound & SG & Lattice Constants (\AA) & $\Delta H_f$ (kcal mol$^{-1}$) & $\Delta H_f$ (eV) & Ref. \\
\hline\hline
57 La   			  &  LaS        & $Fm\bar{3}m$ &  5.788, 5.788, 5.788      & 112.8 & 2.45 & [\onlinecite{LaSstru},\onlinecite{Curta}] \\
			          &  La$_2$S$_3$        & $Pnma$   & 7.66, 4.22, 15.95       & 282.98 & 2.45 & [\onlinecite{La2S3stru},\onlinecite{Curta}] \\
			          &  LaS$_2$        & $Pnma$  & 8.131, 16.34, 4.142       & 162 & 2.34 & [\onlinecite{LaS2stru},\onlinecite{NewBook}] \\ \hline
72 Hf 			 &   HfS$_2$       & $P\bar{3}m1$    &  3.69, 3.69, 6.61      &  &  & [\onlinecite{HfS2stru}] \\
 			      &   HfS$_3$       &  $P2_1/m$   & 5.0923, 3.5952, 8.967    &  &  & [\onlinecite{ZrS3stru}] \\ \hline
73 Ta			  &  TaS$_2$        &   $P6_3/mmc$  &   3.314, 3.314, 12.097     & 84.61 & 1.22 & [\onlinecite{TaS2stru},\onlinecite{Curta}] \\
			          &  TaS$_3$        & $P2_1/m$    &   9.515, 3.3412, 14.912     &  &  & [\onlinecite{TaS3stru},\onlinecite{Curta}] \\ \hline
74 W			  &    WS$_2$      &   $P6_3/mmc$  &  3.1532, 3.1532, 12.323      & 62 & 0.89 & [\onlinecite{WS2stru},\onlinecite{Curta}] \\ \hline
75 Re			  &     ReS$_2$     &  $P\bar{1}$   & 6.455, 6.362, 6.401       &  42.71& 0.93 & [\onlinecite{ReS2stru},\onlinecite{Curta}] \\
			          &     Re$_2$S$_7$     &     &        & 107.91 & 0.52 & [\onlinecite{Curta}] \\ \hline
76 Os			  &   OsS$_2$       &  $Pa\bar{3}$   &  5.6194, 5.6194, 5.6194      & 35.11 & 0.51 & [\onlinecite{OsS2stru},\onlinecite{Curta}] \\ \hline
77 Ir			  &  Ir$_2$S$_3$        &     &        & 59.61 & 0.52 & [\onlinecite{Curta}] \\
			  &  IrS$_2$        & $Pnma$    &  19.791, 3.5673, 5.6242      & 31.81 & 0.46 & [\onlinecite{IrS2stru},\onlinecite{Curta}] \\ \hline
78 Pt		  &    PtS      &  $P4_2/mmc$	& 3.47, 3.47, 6.1          & 19.86 & 0.43 & [\onlinecite{PtSstru},\onlinecite{Curta}] \\
	  &    PtS$_2$      &  	$P\bar{3}m$	& 3.5432, 3.5432, 5.0388           & 26.51 & 0.57 & [\onlinecite{PtS2stru},\onlinecite{Curta}] \\ \hline
79 Au			  &   Au$_2$S      & 	$Pn\bar{3}m$ & 	5.0206, 5.0206, 5.0206        &  &  & [\onlinecite{Au2Sstru},\onlinecite{Curta}] \\ \hline
80 Hg			  &   HgS       & 	$P3_121$ &	4.16, 4.16, 9.54         & 12.74 & 0.28 & [\onlinecite{HgSstru},\onlinecite{Curta}] \\ \hline\hline
\end{tabular}
\end{table*}

\cleardoublepage
\newpage

%\bibliography{Biblio_Corrosion}% Produces the bibliography via BibTeX.


\end{document}